\documentclass[aps,prb,twocolumn,superscriptaddress]{revtex4}

\usepackage{graphicx}
\usepackage{graphics}
\usepackage{amsmath}
\usepackage{subfigure}
\usepackage{color} 
\usepackage{setspace}                           
\usepackage[T1]{fontenc}
\usepackage[all,cmtip]{xy}
\usepackage{tikz-cd}
\begin{document}

\title{Knock-on damage in bilayer graphene: indications for a catalytic pathway}
\author{Jon Zubeltzu}
\email[E-mail: ]{j.zubeltzu@nanogune.eu}
\affiliation{CIC nanoGUNE, Tolosa Hiribidea 76, 20018 Donostia-San Sebastian, Spain}
\author{Andrey Chuvilin}
\affiliation{CIC nanoGUNE, Tolosa Hiribidea 76, 20018 Donostia-San Sebastian, Spain}
\affiliation{Ikerbasque, Basque Foundation for Science, 48011 Bilbao, Spain}
\author{Fabiano Corsetti}
\affiliation{CIC nanoGUNE, Tolosa Hiribidea 76, 20018 Donostia-San Sebastian, Spain}
\author{Amaia Zurutuza}
\affiliation{Graphenea S.A., Tolosa Hiribidea 76, 20018 Donostia-San Sebastian, Spain}
\author{Emilio Artacho}
\affiliation{CIC nanoGUNE, Tolosa Hiribidea 76, 20018 Donostia-San Sebastian, Spain}
\affiliation{Ikerbasque, Basque Foundation for Science, 48011 Bilbao, Spain}
\affiliation{Theory of Condensed Matter, Cavendish Laboratory, University of Cambridge, J. J. Thomson Avenue, Cambridge CB3 0HE, United Kingdom}
\affiliation{Donostia International Physics Center DIPC, Paseo Manuel de Lardizabal 4, 20018 Donostia-San Sebasti\'an, Spain}

\date{\today}

\begin{abstract}
We study by high-resolution transmission electron microscopy the structural response of bilayer graphene to electron irradiation with energies below the knock-on damage threshold of graphene. We observe that one type of divacancy, which we refer to as the butterfly defect, is formed for radiation energies and doses for which no vacancies are formed in clean monolayer graphene. By using first principles calculations based on density-functional theory, we analyze two possible causes related with the presence of a second layer that could explain the observed phenomenon: an increase of the defect stability or a catalytic effect during its creation. For the former, the obtained formation energies of the defect in monolayer and bilayer systems show that the change in stability is negligible. For the latter, \textit{ab initio} molecular dynamics simulations indicate that the threshold energy for direct expulsion does not decrease in bilayer graphene as compared with monolayer graphene, and we demonstrate the possibility of creating divacancies through catalyzed intermediate states below this threshold energy. The estimated cross section agrees with what is observed experimentally. Therefore, we show the possibility of a catalytic pathway for creating vacancies under electron radiation below the expulsion threshold energy.

 \end{abstract}


\maketitle

\section{Introduction}
Nowadays, graphene is one of the most promising and studied materials in the world. The high electronic conductivity and mechanical strength are examples of many singular and desirable properties that this material is characterized by.\cite{Novoselov}$^{\text{-}}$\cite{Lee} However, previous studies have shown\cite{Banhart}$^{\text{-}}$\cite{Telling2} that these singular properties are strongly altered by the presence of defects. Thus, the study of energetics and mechanisms of defect formation, diffusion, and transformation has become an important task in order to control the behavior of graphitic materials: either to maintain their original properties or to change them in a desired way. 

In this respect, high resolution transmission electron microscopy is an ideal tool to carry out this kind of studies, since it provides a controllable impact to the sample by high energy electron flux and, at the same time, the observation of the structural response of the system at the atomic level. Variation of primary electron energy gives a control over the energy transferred to the sample (typically below 20 eV per electronic collision), while variation of the electron flux regulates the rate of transformations.\cite{Santana,Robertson}
In recent years, many types of graphene defects have been analyzed and their energetic and electronic properties have been characterized experimentally and by theoretical simulations.\cite{Palacios}$^{\text{-}}$\cite{Gulans} There is a growing number of experimental studies in which the formation and transformation processes of graphene defects have been observed,\cite{Hashimoto}$^{\text{-}}$\cite{Urita} and the interest in this topic has even increased since the introduction of Cs-corrected microscopes.\cite{Chuvilin,Navarro} Although there has been a substantial theoretical and experimental effort to reveal the mechanisms and key parameters which are responsible for structural transformations in graphene,\cite{Wang}$^{\text{-}}$\cite{Kim} there are still many unanswered questions.
  
\begin{figure}[t!]
 \includegraphics[width=0.5 \textwidth]{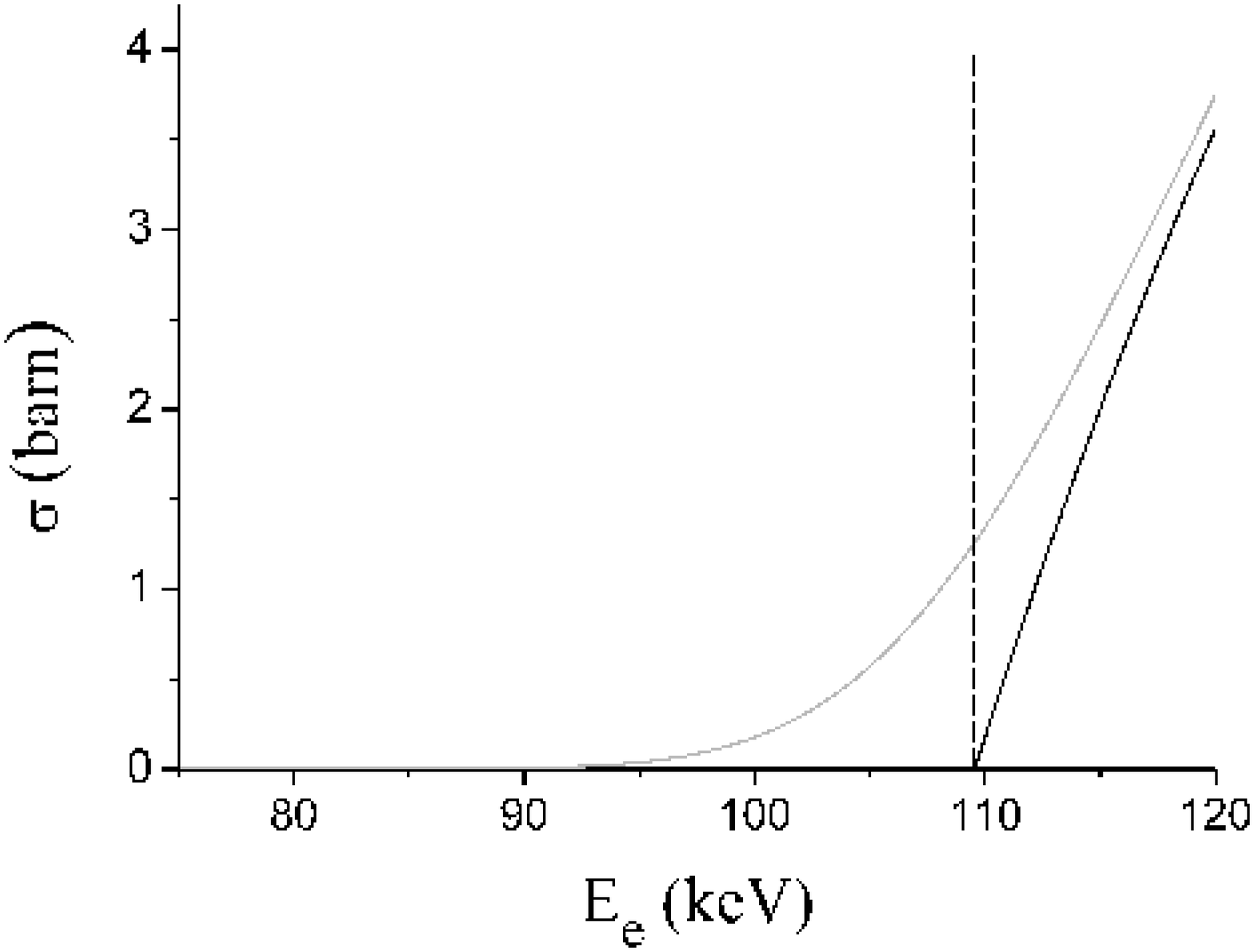}
 \caption{Two different displacement cross sections as a function of the incoming electronic energy. The function obtained by McKinley and Feshbach (dark) predicts a displacement threshold energy of $E_{thr}^{MF}\approx 110$ keV. If the zero-point motion of the carbon atoms in graphene is taken into account (light), the displacement threshold energy is not well defined and the experimental results are much better reproduced. }
 \label{figura1}
 \end{figure}

When radiation energy in a transmission electron microscope is around 100 keV, the formation of vacancies can be observed in a graphene sample.\cite{Banhart,Robertson, Ugeda,Wang,Kotakoski,Meyer2} In order to study the formation mechanism of vacancies by \textit{ab initio} molecular dynamics simulations, usually the classical static lattice approximation is made,\cite{Zobelli,Kotakoski,Wang} where the expulsion threshold energy is defined. Below this energy limit, the likeliness to expel an atom is zero, and thus the creation of vacancies is not possible. McKinley and Feshbach\cite{McKinley} obtained an analytical expression that relates the displacement cross section with the incoming electronic energy within this assumption and predicts a displacement threshold energy of 110 keV for graphene. Recently however, Meyer \textit{et al.}\cite{Meyer2} have shown the importance of the phonon contribution to the displacement cross section: if the zero-point motion is considered, the experimental results are almost perfectly fitted. Figure \ref{figura1} shows the calculated displacement cross section as a function of the radiation energy for both models. When considering the zero-point motion, the tail of the curve descends asymptotically to zero; therefore, the displacement threshold energy is no longer well defined.

As shall be described in subsequent sections, for radiation energies of 80 keV, bilayer graphene shows a substantial increase on the displacement cross section with respect to the monolayer graphene sample. For an electronic dose of the order of $10^{10}$ $e^-$/nm$^2$ the formation of several vacancies is observed in bilayer graphene, while in monolayer graphene the formation of one vacancy is unlikely.\cite{Meyer2}

In this paper, we analyze the possible mechanisms of defect formation under electron radiation in bilayer graphene. We explore and rule out a number of possible explanations for the very high sputtering cross section observed experimentally. We finally propose a new concept of multistep sputtering process and prove its feasibility.

 \section{Methods}

\subsection{Experimental methods}

High resolution transmission electron microscopy (HRTEM) image series are acquired on a Titan 60-300 electron microscope (FEI, Netherlands) equipped with a high brightness electron gun (xFEG), monochromator, imaging Cs corrector, Ultrascan1000 2Kx2K CCD camera, and a GIF Quantum electron energy loss spectrometer (EELS) (Gatan, USA).

The microscope is operated at 80 kV acceleration voltage, the beam is monochromated to about 100 meV energy spread (as measured by full width at half maximum of the zero loss EELS peak), and the image corrector is tuned so that the third-order spherical aberration coefficient is equal to $-$20 $\mu$m. Images are recorded on the pre-GIF camera with an exposition time equal to 1 s.

A post-specimen blanker is used in the experiment, so that the sample is continuously illuminated even between expositions. The dose rate at exposition time is determined from the image intensity, and the dose rate between expositions is linearly interpolated. The total dose is calculated as an integral over time assuming the dose rate as described above.

The image simulations are performed by means of the MUSLI package\cite{Andrey} which is based on the implementation of the fast-fourier-transforms multislice algorithm and assuming neutral atoms. Electron statistics is accounted for in accordance to the experimentally measured dose; the modulation transfer function of the CCD camera is applied on thus simulated images. 

The monolayer graphene samples were grown by chemical vapor deposition (CVD) using 25 $\mu$m copper foil as the catalyst. The monolayer samples were transferred onto Quantifoil Au TEM grids (hole size 2 $\mu$m) using polymethyl methacrylate (PMMA) as the sacrificial polymer layer and ferric chloride as the copper etching agent. In order to prepare the bilayer samples, the transfer process was repeated twice.

\subsection{Theoretical methods}
  To carry out the theoretical calculations, we employ the SIESTA method\cite{Soler} based on density-functional theory (DFT). It is characterized by the use of norm-conserving pseudopotentials\cite{Hamann} and finite-support atomic-like basis-sets. We use the van der Waals density functional (vdW-DF)\cite{Dion} as the exchange-correlation functional in order to take into account van der Waals forces between graphene sheets. Computational parameters have been optimized as follows, in order to achieve a convergence of $1$ meV per atom. We use a double-$\zeta$ polarized (DZP) basis, available in SIESTA's main web page\cite{SIESTA} and a real-space grid with a $100$ Ry mesh cutoff. 
  
To calculate the energy of the pristine and defective systems, we relax the system by the conjugate gradient method,\cite{Hestenes} to within a force tolerance of $0.01$ eV/\AA  \hspace{0.03 cm}. All the relaxation calculations are carried out with no symmetry constraints. We use a large enough supercell to contain our defect and sufficiently reduce the finite-size effects of the calculations. The employed rectangular supercell contains 384 atoms and the edges are defined as:
\begin{equation}
\dbinom{\textbf{L}_x}{\textbf{L}_y}=\begin{pmatrix} 8 & 0 \\ -6 & 12 \end{pmatrix}\dbinom{\textbf{a}_1}{\textbf{a}_2},
\end{equation}
where $\textbf{a}_1$ and $\textbf{a}_2$ are the primitive lattice vectors of graphene defined as in Shallcross \textit{et al.}\cite{Shallcross} The $y$ edge of the supercell is larger because the extension of the defect in that direction is larger. For the \textbf{k}-point sampling of the Brillouin zone a ($3\times2\times1$) Monkhorst-Pack matrix\cite{Monkhorst} is chosen.

Although one layer is rotated respect to the other in the bilayer graphene sample, we will assume that both sheets are always in a parallel orientation. The rotation of one of the layers in our simulation box would break the periodicity in our supercell and a much bigger cell would be needed to carry out the calculations. To reproduce the different local stackings that are formed in the sample, we translate one of the layers. 

In Table \ref{tabla1} we show the obtained values for the inter planar distance between two graphene layers $c$ and the energy difference per atom between a bilayer graphene system with an AA and AB stacking $\Delta E_{AA/AB}$. In order to check the reliability of the calculations, we add the same magnitudes obtained by Birowska \textit{et al.}\cite{Birowska} with the same exchange-correlation functional (vdW-DF) and the experimental value for the interlayer distance for graphite.\cite{Zhao} If we compare the data shown in Table \ref{tabla1} we conclude that the obtained interlayer distance is close to its experimental value, and that the difference in $\Delta E_{AA/AB}$ is within the error of $1$ meV per atom. 

In the kinetic analysis, we will make two justified assumptions for simplicity: firstly, as the rate for a thermally activated process is proportional to $e^{-{\frac{\Delta E}{kT}}}$, and in our case, the activation barriers $\Delta E$ are of the order of 1 eV, we may neglect thermally activated processes and consider activation by electron collision only. Secondly, the time interval between a scattering event involving a given atom and a second scattering event in its neighborhood is greater than the relaxation time of the system, thus the processes activated by electronic radiation will be treated as being caused by singular scattering events and the system will always remain relaxed between these scattering events. 

For simulating the kinetic process after the collision, we suppose that the electrons are coming from $z =-\infty$ having a velocity parallel and positive in the $z$ axis. The graphene layers remain perpendicularly oriented to the electron beam. 

For describing the evolution of the system after an electron scattering event, we use \textit{ab initio} molecular dynamics (AIMD): we divide the time in $1$ fs timesteps and for each one the forces are calculated on each atom using the DFT method discussed above. The Eqs. of motion are then solved by Verlet integration.\cite{Verlet} At time $t=0$ the system remains relaxed and all the atoms are at rest, thus neither temperature nor zero-point motion contributions are taken into account. Therefore, within this model, we can define the expulsion threshold energy, which is the minimum energy needed to expel an atom from the system. 

To simulate the collision of the electron with the atom, a certain velocity is given to one atom from the sample. In this case, as the initial states from which the AIMD simulations are initiated do not contain large defects, we use a smaller supercell to reduce the computational cost. The dimensions of the supercell are: 
\begin{equation}
\dbinom{\textbf{L}_x}{\textbf{L}_y}=\begin{pmatrix} 7 & 0 \\ -5 & 10 \end{pmatrix}\dbinom{\textbf{a}_1}{\textbf{a}_2},
\end{equation}
and it contains 280 atoms. For the $\bf{k}$-point sampling a ($3\times3\times1$) Monkhorst-Pack matrix is employed.

 \begin{table}[t]
\caption{Interlayer distance and energy difference per atom between AA and AB stackings in bilayer graphene and graphite.}
\centering
\begin{ruledtabular}
    \begin{tabular}{c | c @{\hspace{0.5 cm}} c@{\hspace{0.5 cm}}  c }
      & System & $c$ ($\text{\AA}$) & $\Delta E_{AA/AB}$ $(\frac{\text{meV}}{\text{atom}})$ \\ \hline
    This study & Bilayer & $3.343$ & 8.5  \\ 
    Birowska \textit{et al.}\cite{Birowska} & Bilayer  & 3.349 & 7.5  \\ 
    Expt.\cite{Zhao}& Graphite  & 3.356 &   \\
    \end{tabular}
 \end{ruledtabular}
    \label{tabla1}
    \end{table}

Considering the electron as a relativistic particle and the atom as a classical one which remains at rest, we can obtain an analytical expression for the maximum kinetic energy that is transferred to the latter in a pure elastic head-on collision:\cite{Zobelli}
\begin{equation}
 T_{max}=\frac{2ME(E+2mc^2)}{(M+m)^2c^2+2ME}\hspace{1.0 mm} ,
 \label{eq.2}
 \end{equation}
  where $T_{max}$ is the maximum kinetic energy of the atom along the same direction of the incident electron, $M$ and $m$ are the masses of the atom and electron respectively, $c$ is the speed of light, and $E$ is the energy of the electron. If the atom is emitted in another direction, the maximum obtainable kinetic energy becomes

\begin{figure*}[t!]
\centering
\subfigure[\ ]{\includegraphics[width=0.496 \textwidth]{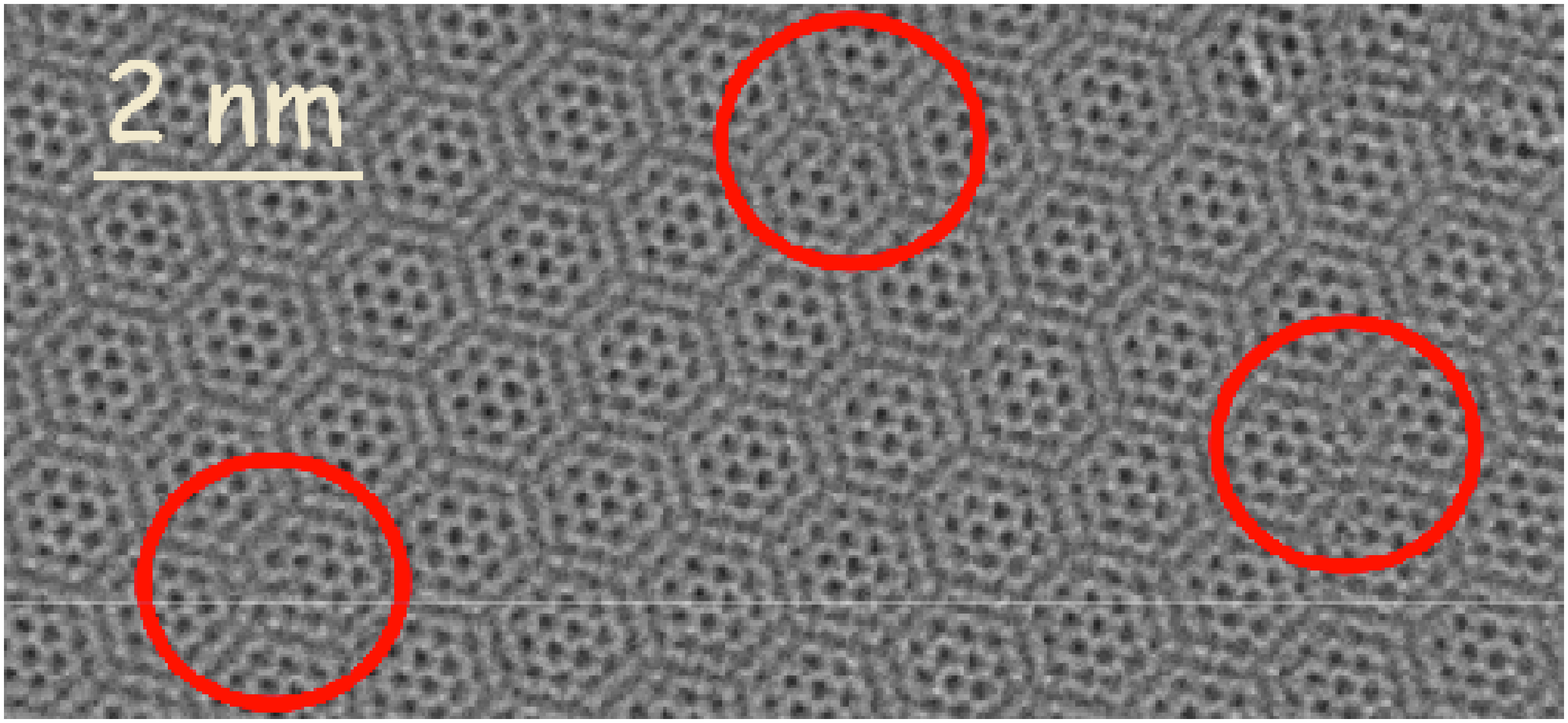}}\hspace{0.0 mm}
\subfigure[\ ]{\includegraphics[width=0.496 \textwidth]{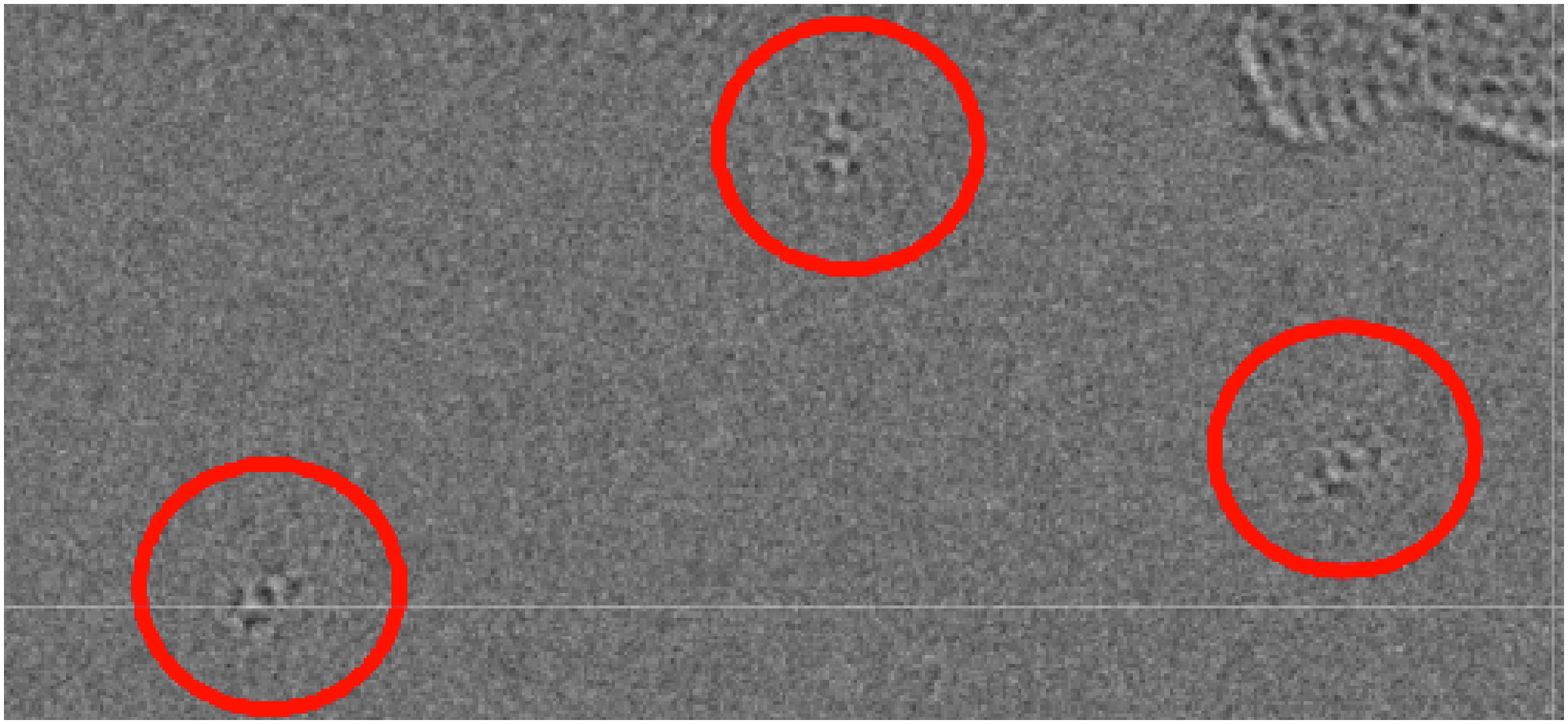}}\\
\subfigure[\ ]{\includegraphics[width=0.245 \textwidth]{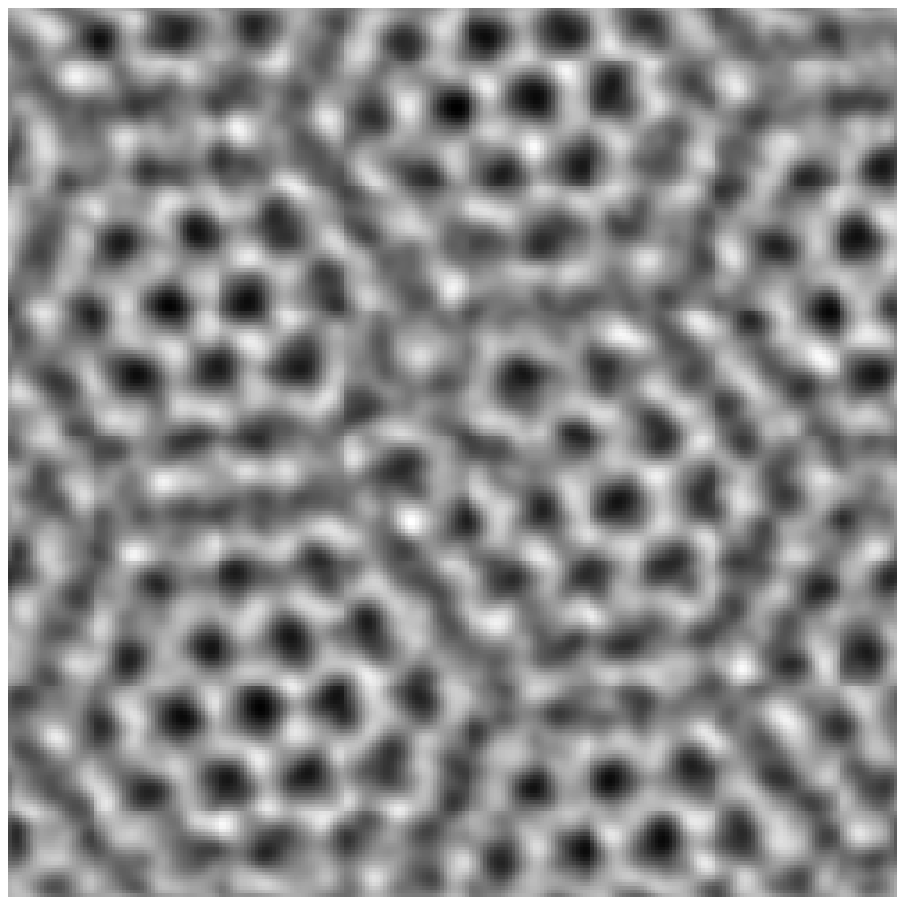}}
\subfigure[\ ]{\includegraphics[width=0.245 \textwidth]{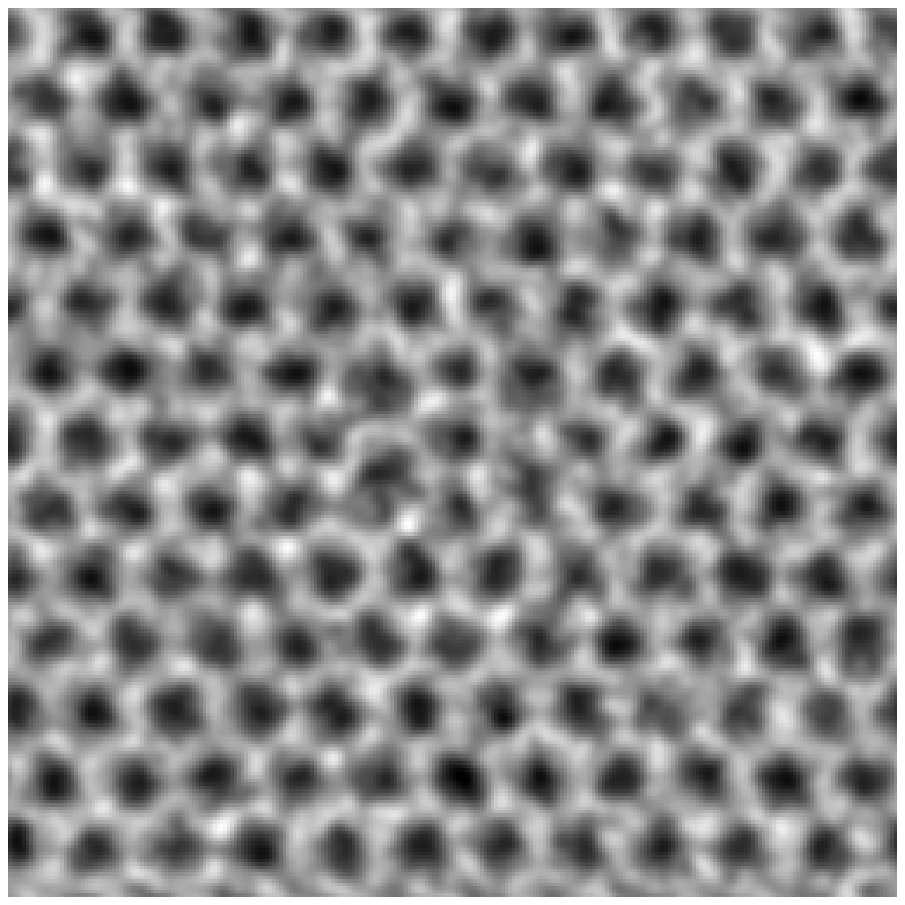}}
\subfigure[\ ]{\includegraphics[width=0.245 \textwidth]{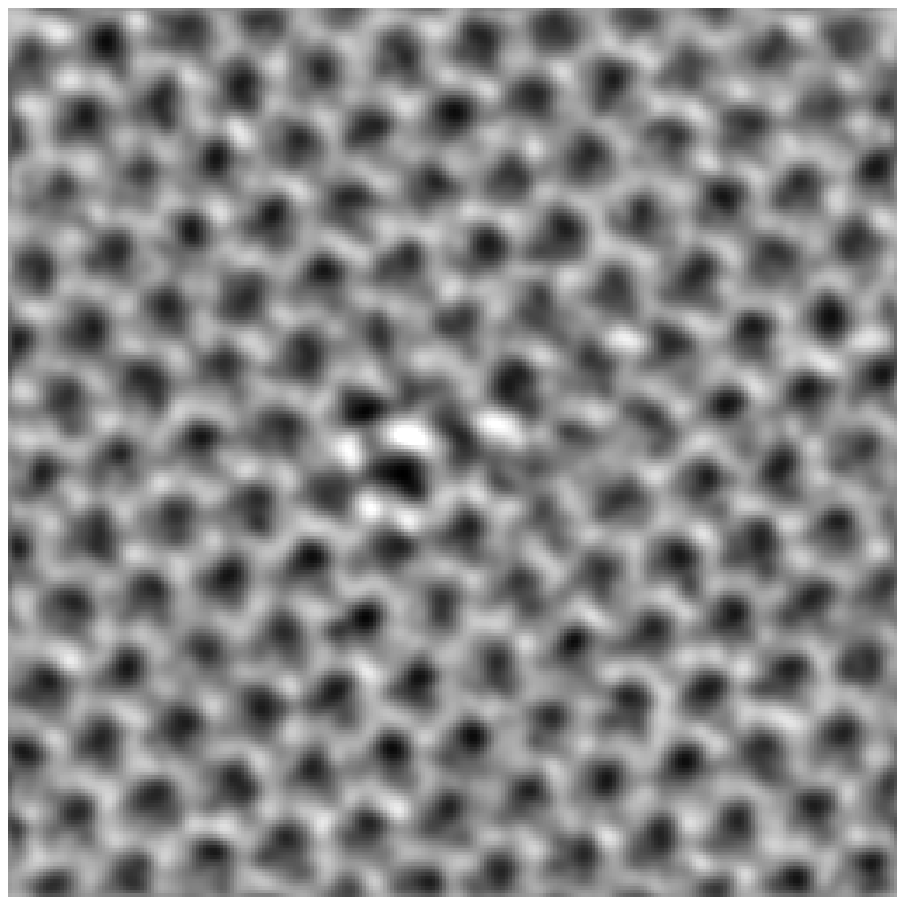}} 
\subfigure[\ ]{\includegraphics[width=0.245 \textwidth]{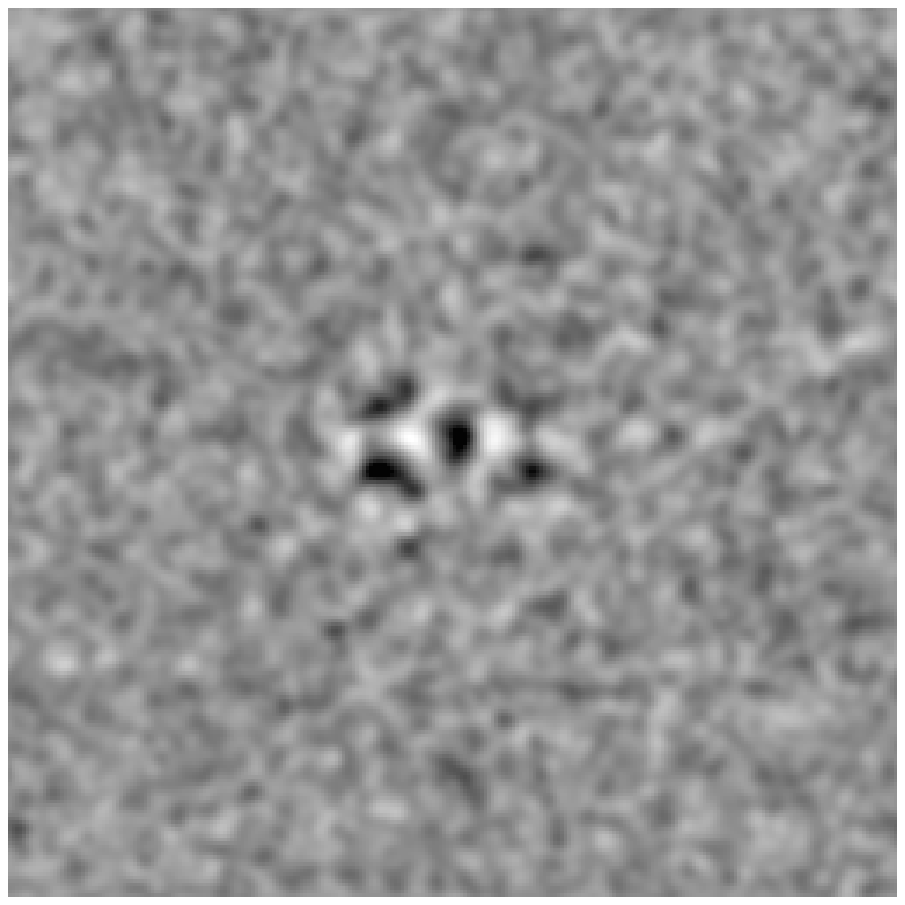}}\hspace{0.0 mm}
\label{figura2}
\caption{Experimental images obtained by HRTEM on bilayer graphene. (a) Overview image showing Moir\'e pattern. In the regions marked by the circles the distortions of the pattern are visible. (b) The same area, but the honeycomb lattices of both layers are removed by Fourier filtering. Characteristic patterns in the shape of double dumbbells appear inside the circles indicating the same type of lattice distortion in all three places. (c) Enlarged image of one of the distorted Moir\'e pattern areas. (d)$-$(f) Fourier filtered images of enlarged area with the second  (d), first (e) and both graphene layers filtered out (f). The image of the first layer only (d) can be directly interpreted in terms of atom positions and reveals a  V$_2$(5555-6-7777) butterfly defect formation in this layer. Image of thesecond layer (e) cannot be interpreted directly and needs simulations in order to find its origin.}
 \label{figura2}
\end{figure*}
  \begin{equation}
 T_{max}(\theta)=T_{max}{\cos^2{\theta}},
 \label{eq.3}
 \end{equation}
  where $\theta$ is the emission angle and is defined by the angle between the direction of the incident electron and emitted atom. Therefore, by using Eqs. (\ref{eq.2}) and (\ref{eq.3}) we obtain the maximum kinetic energy that an atom from the sample can achieve for a certain emission angle $T_{max}(\theta)$ from the energy $E$ of the electrons in HRTEM. The kinetic energy given to the atom at the initial state of the AIMD simulations will be equal to or smaller than the maximum kinetic energy achievable by the atom because of the collision with the electron. 

To estimate the defect population in the sample, we will assume that the rate of a given reversible reaction activated by electron-atom collisions,
\begin{equation}
\mbox{A}\longleftrightarrow \mbox{B},
 \label{eq.6}
 \end{equation}
 follows a first order rate law:
 \begin{equation}
 -\dfrac{d[\mbox{A}]}{dt}=k_f[\mbox{A}]-k_b[\mbox{B}],
 \label{eq.7}
 \end{equation}
 where $[\mbox{A}]$ and $[\mbox{B}]$ are the time-dependent concentrations of A and B species respectively, while $k_f$ and $k_b$ are the rate constants for the forward and backward reactions. If the reactions are activated only by electron collisions, the rate constant is given by\cite{Santana}
 \begin{equation}
k=\sigma j,
 \label{eq.8}
 \end{equation}
 where, $\sigma$ is the cross section related with the process, and $j$ is the electronic dose rate. Assuming that the initial concentration of B is zero, $[\text{B}]_0=0$, the following condition must be fulfilled:
 \begin{equation}
 [\mbox{A}]+[\mbox{B}]=[\mbox{A}]_0.
 \label{eq.9}
 \end{equation}
 The solutions of Eqs. (\ref{eq.7}) and (\ref{eq.9}) are
 \begin{align} 
 [\text{A}] & = ([\mbox{A}]_0-[\mbox{A}]_e)e^{-(k_f+k_b)t}+[\mbox{A}]_e, \label{eq.20}\\ 
 [\text{B}]&=(1-e^{-(k_f+k_b)t})[\mbox{B}]_e, \label{eq.21}
\end{align}
where $[\text{A}]_e$ and $[\text{B}]_e$ are the equilibrium concentrations for each species. In equilibrium, the reaction rate must be zero, $d[\mbox{A}]/dt=0$, and thus from Eq. (\ref{eq.7}) we can calculate the equilibrium relative concentration between A and B:
 \begin{equation}
\dfrac{[\mbox{A}]_e}{[\mbox{B}]_e}=\dfrac{\sigma_b}{\sigma_f},
 \label{eq.10}
 \end{equation}
where $\sigma_b$ and $\sigma_f$ are the cross sections related with the backward and forward reactions respectively. The possibility of using Eq. (\ref{eq.10}) to analyze our results is determined by the reaction velocity with which the system is approximated to the equilibrium state, given by the exponent in Eqs. (\ref{eq.20}) and (\ref{eq.21}): $k_f+k_b$. In our case, the radiation exposure time of the sample is longer than $1/(k_f+k_b)$, thus the use of Eq. (\ref{eq.10}) is well justified.

In order to estimate the cross section related with a given scattering event, we will use the expression of the impact parameter in a Coulomb scattering for a semiclassical relativistic electron:\cite{Matzdorf}
\begin{equation}
b=\frac{Ze^2\tan\theta}{4\pi\varepsilon_0m\gamma v^2},
 \label{eq.4}
 \end{equation}
where $Z$ is the atomic number of the target atom, $e$ is the electron charge, $\varepsilon_0$ is the vacuum permittivity, $\gamma$ is the Lorentz factor and $v$ is the velocity of the electron. This last expression is obtained by assuming that the target atom is much heavier than the electron ($M\gg m$) and large bombarding energies. The angle $\theta$ is obtained from Eq. (\ref{eq.3}), where $T_{max}(\theta)$ in this case, is taken as the activation energy for a given process. Once the impact parameter is known, the cross section is given by:
\begin{equation}
\sigma=\pi b^2.
 \label{eq.5}
 \end{equation}

In some case, the activation energies that we will use are thermal barriers, i.e., the minimum energy required by the system in order to activate a process. However, we are assuming that these barriers are isotropic in the \textit{xy} plane, and once the activation energy is obtained by the scattered atom the process will always be initiated. In addition, as the atoms remain frozen and the energy is given only to the  scattered atom, the barrier that this atom will have to overcome is always greater than the thermal one, which is not taken into account in the previous equations. Therefore, the cross sections and consequently defect concentration that we will estimate in the results will always be overestimated.
  
\section{Results}

\subsection{Experimental results}
 
 \begin{figure*}[t!]
 \subfigure[\ ]{\includegraphics[width=0.3 \textwidth]{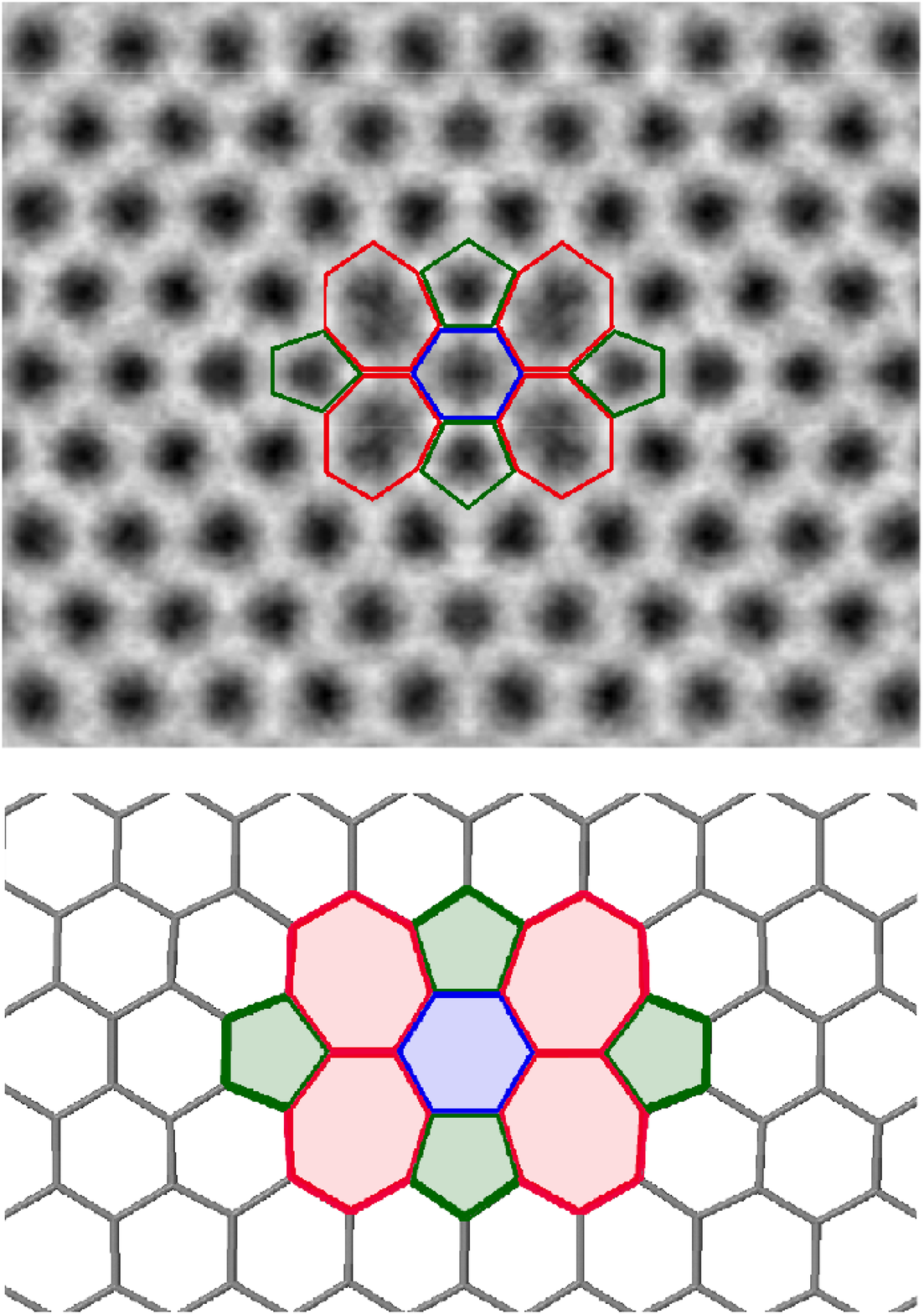}}
 \label{figura3a}
 \vspace{.20in}
 \subfigure[\ ]{\includegraphics[width=0.65 \textwidth]{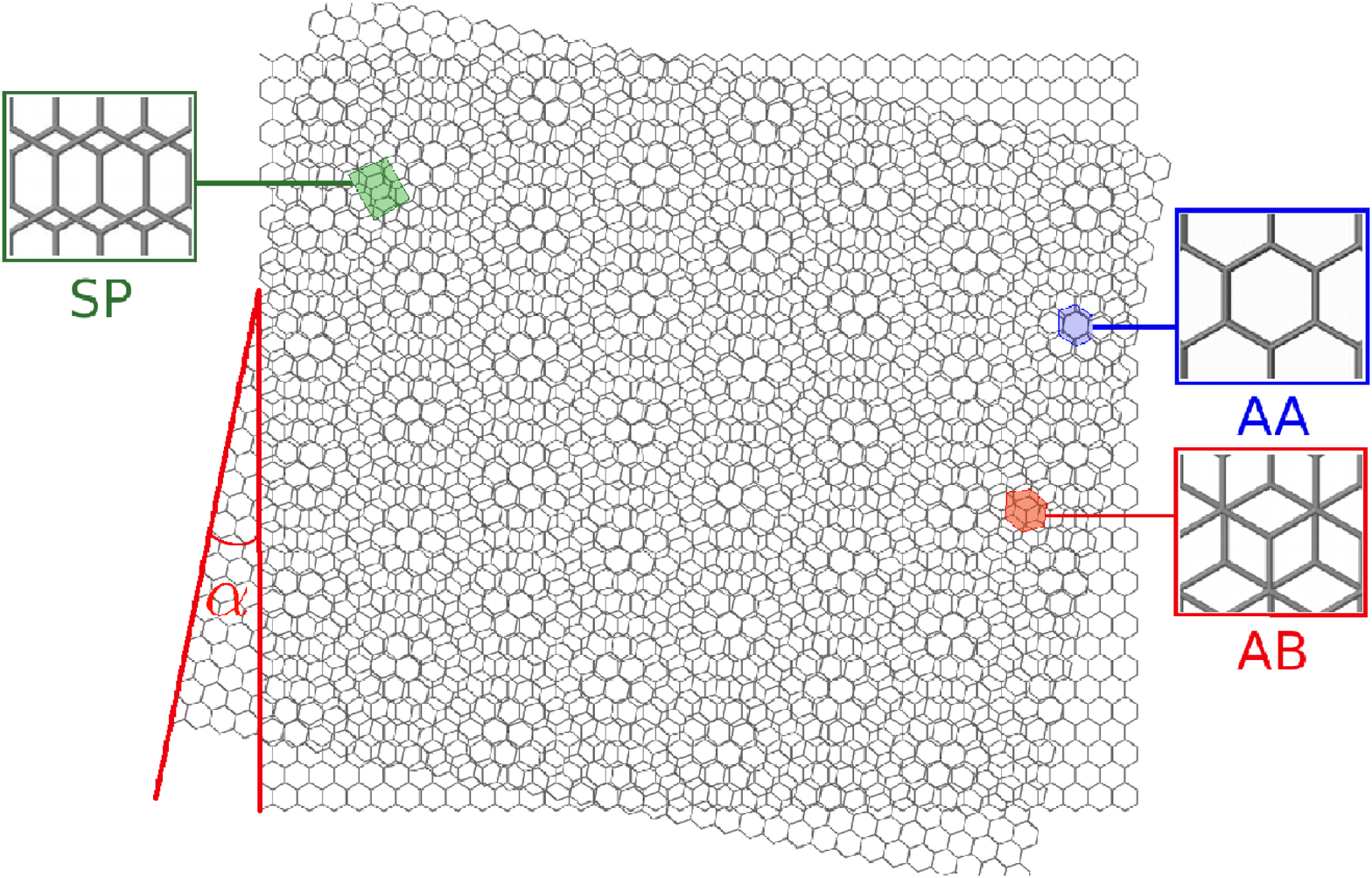}}
 \caption{(a) Experimental image averaged over all observed butterfly defects clearly showing its structure and a corresponding drawing of the defect. Four heptagonal (red), four pentagonal (green) and one rotated central hexagonal (blue) rings are formed instead of the original hexagonal lattice. (b) Atomic model of the graphene layers used for image simulations. One of the layers is rotated an angle $\alpha = 11.2^{\circ}$. Different local stackings are formed in different small areas within the large hexagons.}
 \label{figura3}
\end{figure*}
 \begin{figure*}[t!]
\hspace{0.3mm}
 \subfigure[\ ]{\includegraphics[width=0.243 \textwidth]{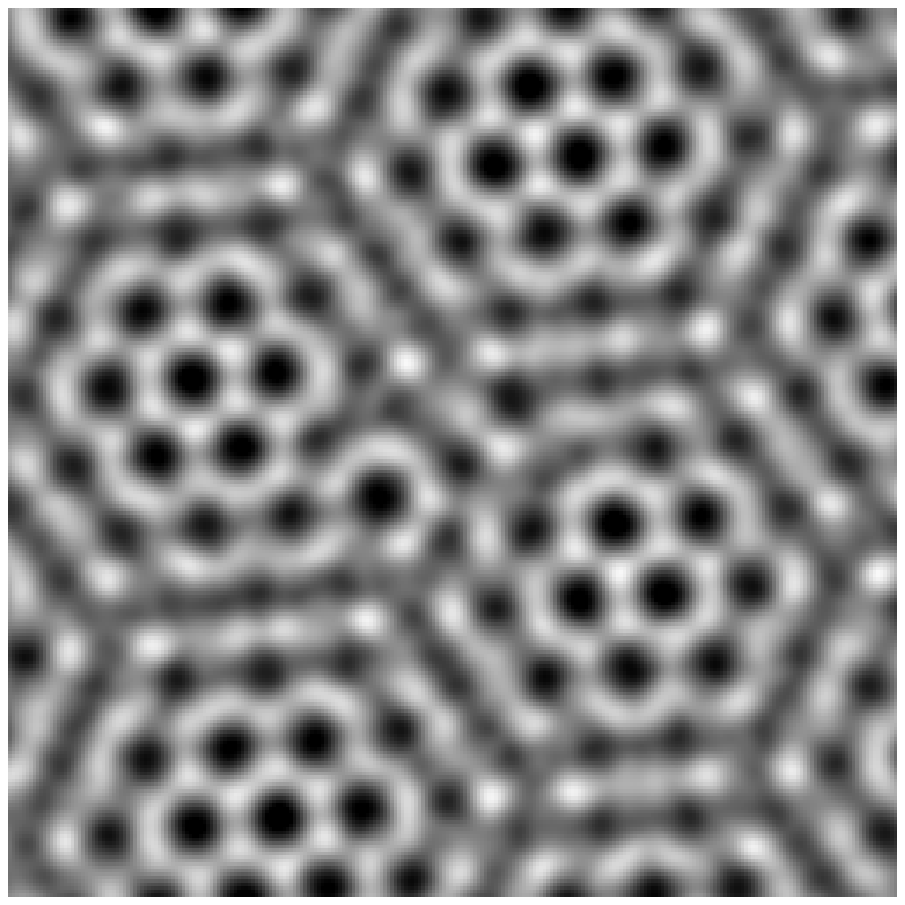}}
 \subfigure[\ ]{\includegraphics[width=0.243 \textwidth]{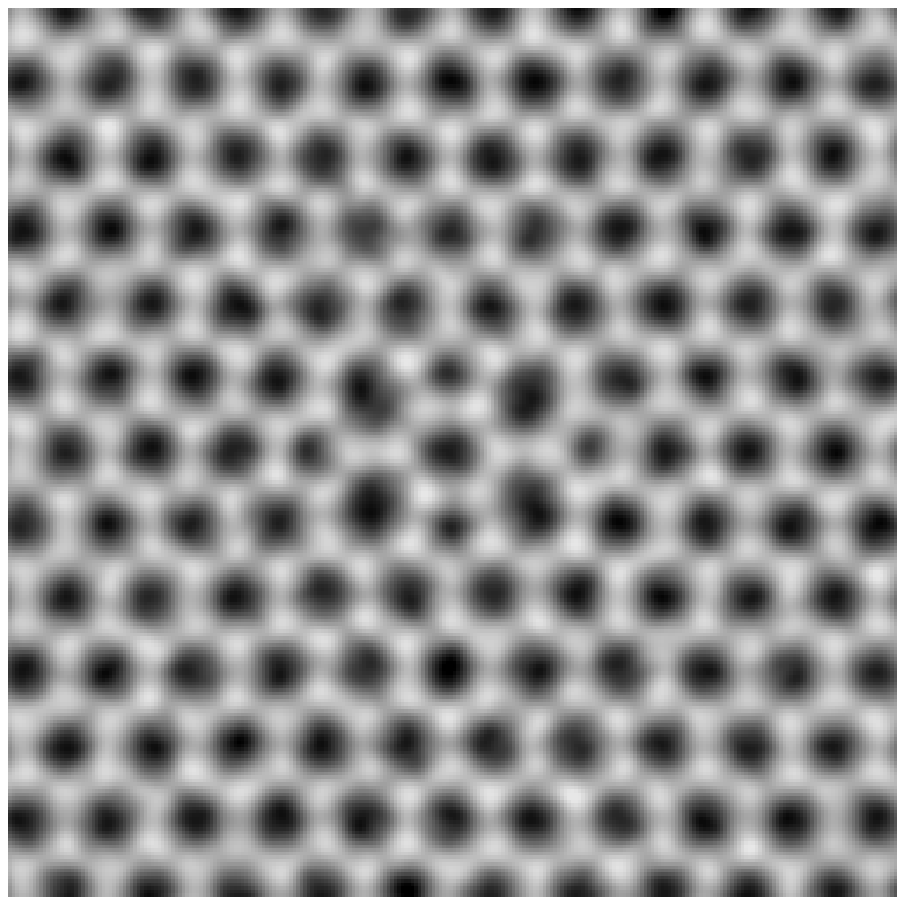}}
  \subfigure[\ ]{\includegraphics[width=0.243 \textwidth]{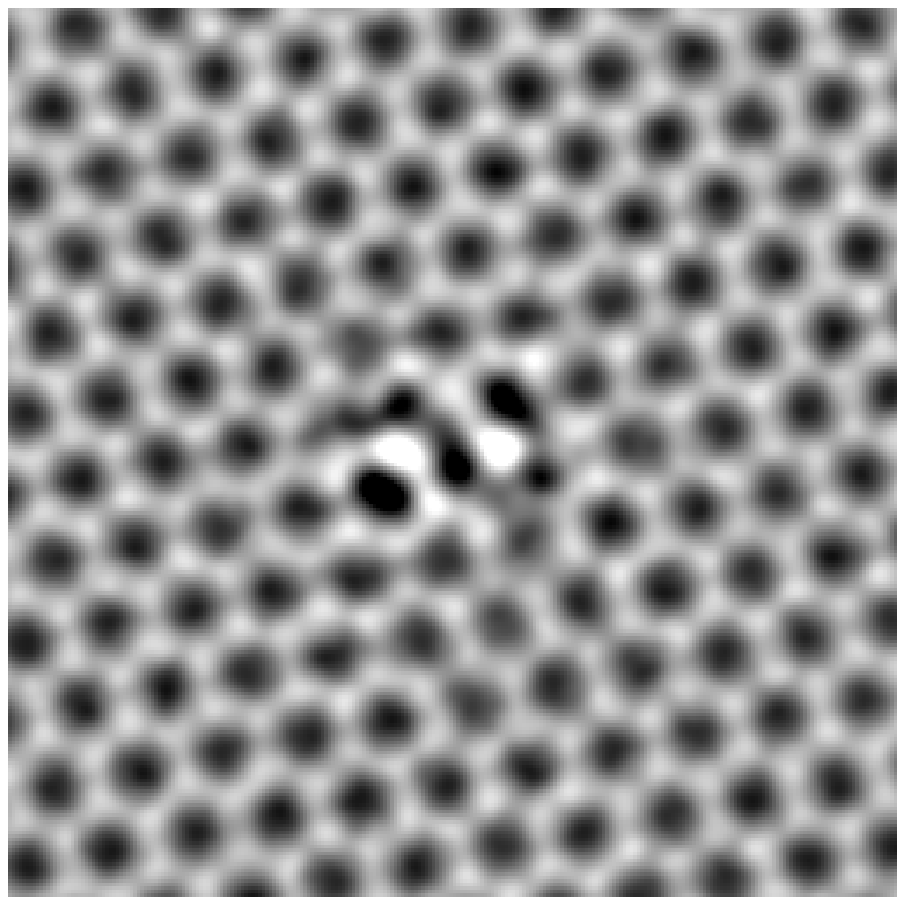}} 
   \subfigure[\ ]{\includegraphics[width=0.243 \textwidth]{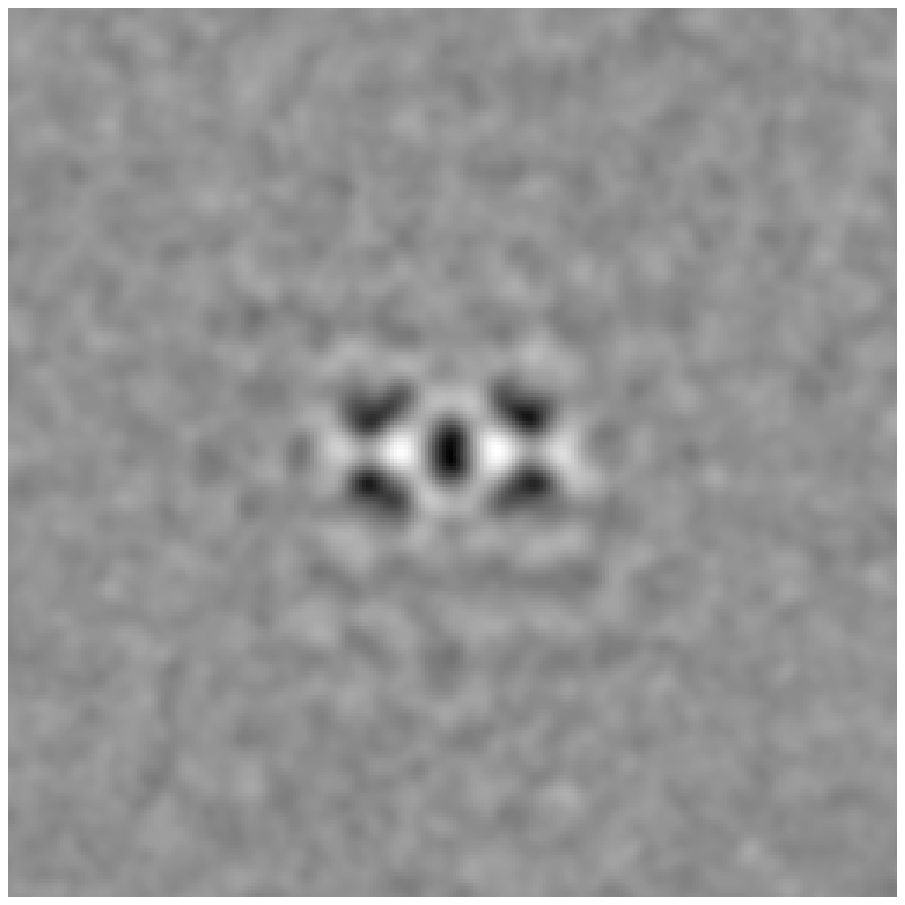}}\hspace{0.3 mm}

 \caption{Central part of the simulated image shown at the same scale and with the same processing as experimental images in Figs. \ref{figura2}(c)$-$\ref{figura2}(f). The butterfly defect is located in the first layer and the second one is kept pristine. The feature similar to Fig. \ref{figura2}(e) is observed on the simulated second layer (c) comprising pristine graphene. It can be thus concluded that this feature in the second layer is an artifact of Fourier filtering.}
 \label{figura4}
\end{figure*}

In Fig. \ref{figura2} we observe a bilayer graphene sample with a rotation angle between layers of 11.2$^\circ$. A characteristic hexagonal Moir\'e pattern is observed in HRTEM images due to this rotational misfit. After extended observation time we start to observe distortions on the Moir\'e figures, which are attributed to radiation generated defects. Fourier filtering of lattice patterns of one and the other layer clearly reveals that we observe V$_2$(5555-6-7777) type divacancies, in one and the same layer always. From HRTEM images only it is impossible to determine whether the layer containing the defects is an upper or lower one with respect to electron beam propagation direction (from theory we can, as shown later). The filtered image of the second layer represents an irregular pattern at the position of the defect, which is not possible to interpret directly. Removal of both lattices from the image produces a characteristic signature of a V$_2$(5555-6-7777) defect in the shape of dumbbell. Figure \ref{figura2} shows all the observations described above.

\begin{figure*}[t!]
 \subfigure[\ ]{\includegraphics[width=0.4 \textwidth]{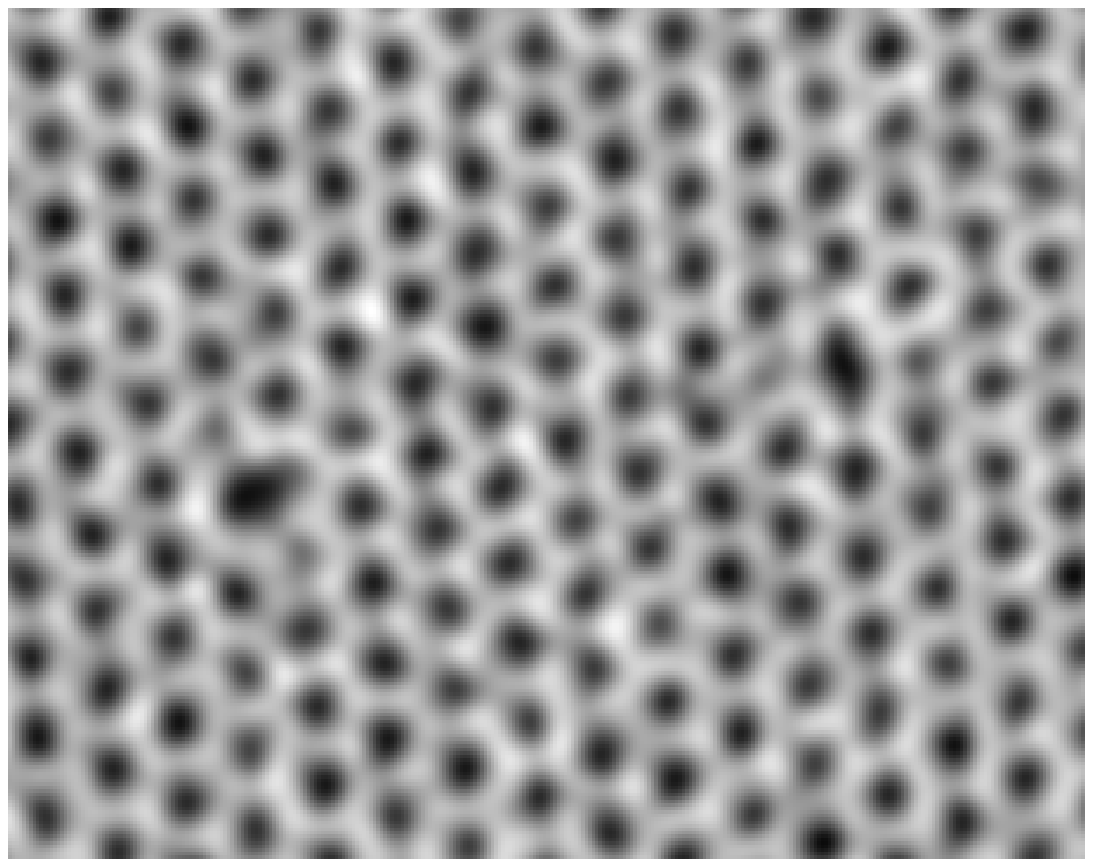}}\hspace{2.0 mm}
 \subfigure[\ ]{\includegraphics[width=0.4 \textwidth]{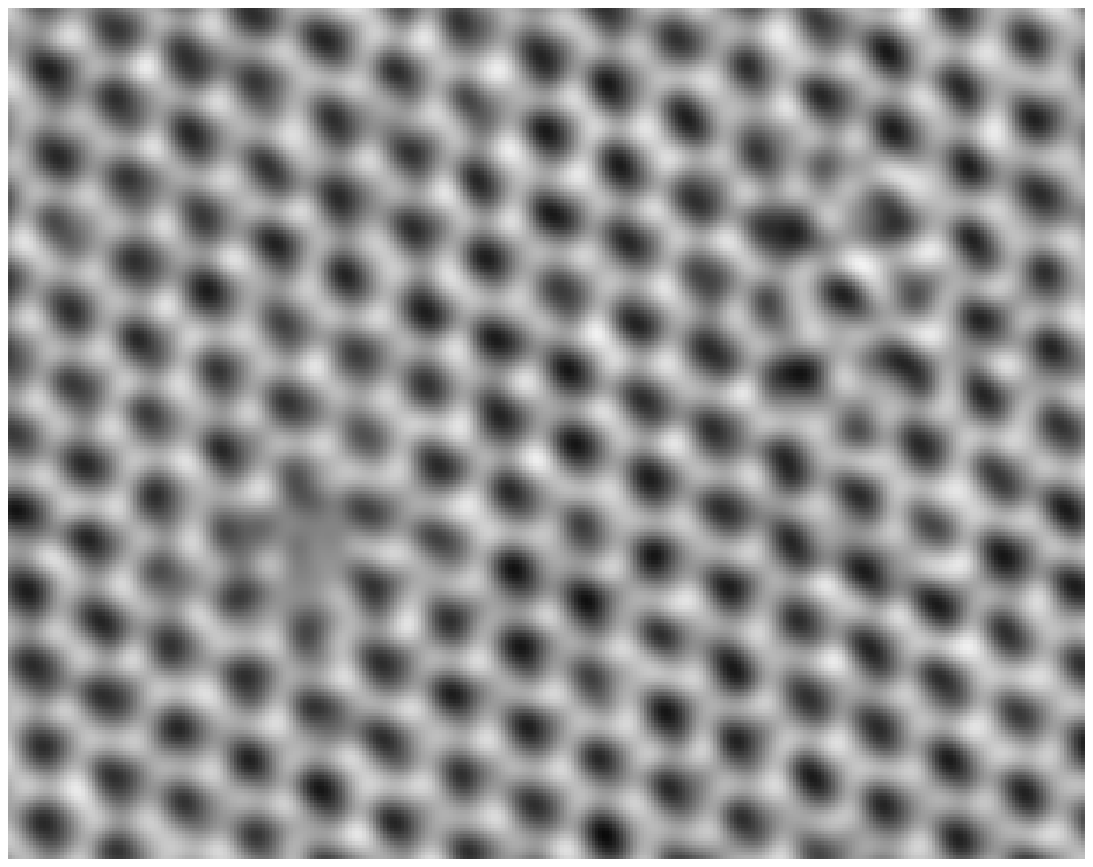}}
 \caption{Enlarged experimental HRTEM images obtained during experiment. The second honeycomb lattice is filtered out. (a) Two V$_2$(5-8-5) divacancies are distinguished in the first. (b) After 7.5 $e$/nm$^2$ of electronic dose, one of the divacancies (left) is in the process of transformation while the other (right) is converted into a butterfly defect.}
\label{figura5}
\end{figure*}

The simulation of the observed defect confirms that the formed defects are V$_2$(5555-6-7777) divacancies, or for simplicity, butterfly defects (Fig. \ref{figura3}). They are characterized by the formation of four heptagonal, four pentagonal, and one rotated central hexagonal carbon rings. The simulation of the formed Moir\'e pattern is shown in Fig. \ref{figura3}, where we can clearly distinguish three characteristic zones depending on the different local stackings: an AA stacking at the center of the hexagons, a local AB stacking at the corners, and a saddle point (SP) stacking at the edges. This last one can be obtained if one layer is translated a half-bond distance through a bond direction from the AB stacking

If we pay attention to the location of the defects in Fig. \ref{figura2}, we observe that they are stabilized close to the edges (SP) and corners (AB) of the large hexagons of the Moir\'e pattern in all cases. This last observation suggests that the stacking influences the stabilization of the defects.

The HRTEM image simulation of the bilayer rotated by 11.2$^\circ$ with V$_2$(5555-6-7777) defect presented in one of the layers reproduces exactly all experimentally observed features (Fig. \ref{figura4}): the Moir\'e distortion, the V$_2$(5555-6-7777) image in one of the layers while applying the same Fourier filter, the disordered structure in the second layer, and the dumbbell signature when both lattices are filtered out. On the basis of this analysis we can conclude that we do really see divacancy generation in one of the layers, and the disordered structure observed in the second layer is an artifact of Fourier filtration.

During image series acquisition, it is observed that V$_2$(5-8-5) (Fig. \ref{figura5}) and V$_2$(555-777)\cite{Chuvilin2} divacancies are formed before they are converted into butterfly defects.  The evolution V$_2$(5-8-5)$\longrightarrow$V$_2$(5555-6-7777) can be understood by two Stone-Wales transformations:\cite{Kotakoski,Wang} if one of the bonds from V$_2$(5-8-5) is rotated $90^{\circ}$, a V$_2$(555-777) divacancy is formed and, if once again a second bond is rotated, the butterfly defect is obtained. 

We monitor the total deposited dose from the pristine bilayer until three butterfly defects are observed in the field of view with the area of 52 nm$^2$. The total dose accumulated during this observation is 1.3$\times$10$^{10}e^-$/nm$^2$. The cross section of C atom sputtering calculated from this data is $\sigma$=1.2 mb, which is by at least two orders of magnitude higher than the estimation of the low limit of sputtering cross section for a single layer.\cite{Meyer2} In combination with the fact that the vacancies were only created in one layer, our observation points to a strong synergetic influence of the second layer on the sputtering process. Hereafter we evaluate possible mechanisms which may contribute to this synergy.

\subsection{Theoretical results}

We consider two possible causes that could be behind this phenomenon: an increase of the stability of the defect from monolayer to bilayer graphene, or a catalytic effect of the second layer during the creation process of the divacancy. 

\subsubsection{Stability}
In order to measure the stability of the butterfly defect in monolayer and bilayer graphene we calculate its formation energy, defined as
\begin{equation}
E_f=E_{def}-E_{bulk}-\Delta n\mu_c,
\end{equation}
where $E_{def}$ is the total energy of the $N$-atom supercell with a single defect, $E_{bulk}$ is the total energy of the same supercell containing perfect crystal, $\Delta n$ is the required change in atom number to create the defect, and $\mu_c$ is the chemical potential of carbon in the pristine configuration:
\begin{equation}
\mu_c=\frac{E_{bulk}}{N}.
\end{equation}
The lower the value is of the formation energy of the defect, the higher is its stability. For the butterfly defect in monolayer graphene we obtain

\begin{figure}[b!]
 \includegraphics[width=0.5\textwidth]{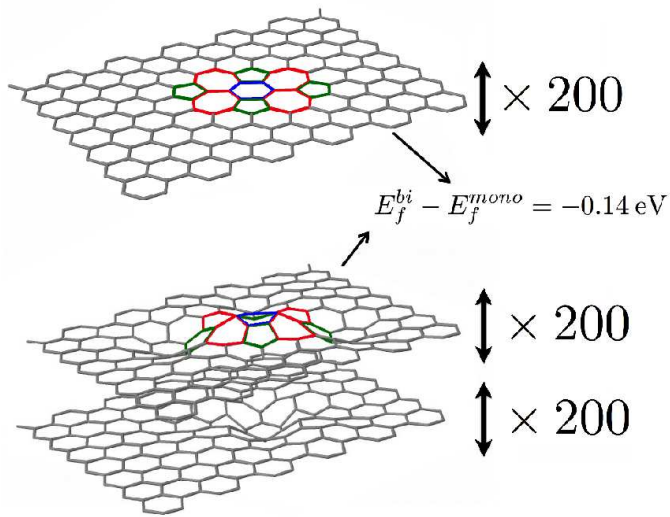}
 \caption{Calculated atomic displacements multiplied by a factor of 200 in the perpendicular direction of the graphene sheets made by the presence of the butterfly defect in monolayer and bilayer graphene. The formation energy difference of the defect between both systems is negligible in this context.}
 \label{figura6}
 \end{figure}

\begin{equation}
E_f^{mono}=7.08 \: \text{eV},
\end{equation}
while in the bilayer case
\begin{equation}
E_f^{bi}=6.94 \: \text{eV}.
\end{equation}
The stability of the butterfly defect increases from monolayer to bilayer graphene, but the energetic change is very small: 
\begin{equation}
\Delta E_{f}=E_f^{bi}-E_f^{mono}=-0.14 \: \text{eV}.
\end{equation}
 
Indeed, there is only a 2$\%$ decrease of the formation energy, not sufficient to explain the big change of the displacement cross section that is observed experimentally. We conclude that the difference in energetics is not the cause of the observed phenomenon and the mechanism for the formation of the butterfly defect is different in monolayer and bilayer graphene.

The analysis of the atomic displacements that occur in both systems supports the previous conclusion: Fig. \ref{figura6} shows the deformations that take place in monolayer and bilayer systems in the perpendicular direction multiplied by a factor of 200. For monolayer graphene, there are no appreciable displacements, while in the bilayer case we can observe that in the upper layer (the one closer to the beam source as is shown below) the central hexagon of the butterfly defect ascends and another hexagon from the lower layer (farer from the beam source) descends. However, the maximum displacements are of the order of $10^{-3}$ \AA, consistent with the very small energetic changes. 

We calculate the formation energies of the V$_2$(5-8-5) and V$_2$(555-777) divacancies in monolayer and bilayer graphene. Table \ref{tabla2} summarizes the obtained results. The formation energies of each kind of divacancy change very little from the monolyaer system to the bilayer one. Therefore, the previous conclusion is confirmed: the energetic analysis does not explain the observed phenomenon. Looking at the results in Table \ref{tabla2}, one would expect that the most stable divacancy is the V$_2$(555-777). However, this is in clear contrast with the experimental observation: during the electronic radiation the created divacancies form the different structures V$_2$(5-8-5) (as seen in Fig. \ref{figura5}) and V$_2$(555-777),\cite{Chuvilin2} but they all finally evolve into the butterfly structure, which then remains stable. The reason for the discrepancy with the results in Table \ref{tabla2} remains unknown, being possibly related to dynamical or entropic effects.

\subsubsection{Kinetics: one-step sputtering process}

\begin{table}[t!]
\caption{The formation energies of three types of divacancy in monolayer and bilayer graphene}
\centering
\begin{ruledtabular}
    \begin{tabular}{ c | c @{\hspace{0.5 cm}} c@{\hspace{0.5 cm}}  @{\hspace{0.5 cm}}}
     &$E_{f}^{mono}($eV$)$&$E_{f}^{bi}($eV$)$ \\ \hline
    V$_2$(5-8-5) & 7.28 & 7.32   \\
    V$_2$(555-777) & 6.74 & 6.64   \\ 
    V$_2$(5555-6-7777)& 7.08 & 6.94   \\ 
    \end{tabular}
\end{ruledtabular}
\label{tabla2}
\end{table}

We calculate the amount of energy needed to remove an atom in the monolayer system using AIMD and verify if for lower energetic values it is possible to expel an atom in bilayer graphene. We find that the expulsion threshold energy for monolayer graphene is $22$ eV, in good agreement with previous studies.\cite{Zobelli,Kotakoski2} In the case of the bilayer system, we employ a sample in AB stacking. Remembering that the electron beam comes from above, we can distinguish three types of atoms depending on their configuration: An atom located in the lower layer (A), one that  is situated in the upper layer and centered with respect to a lower carbon hexagon (B), and one that is located in the upper layer but is directly on top of a lower atom (C). Table \ref{tabla3} shows the expulsion threshold energy for each type and the corresponding electronic energy obtained by Eq. (\ref{eq.2}). Our results show that the energy needed to remove an atom from the bilayer system is equal or greater than that needed for the monolayer.

Since in the case of monolayer graphene it has been shown that the phonon contribution plays an important role in the theoretical explanation of the observed experimental displacement cross section,\cite{Meyer2} we analyze possible effects generated by lattice vibrations in bilayer graphene. By using the Eqs. in Meyer \textit{et al.}\cite{Meyer2} we obtain that the original graphene's Debye temperature perpendicular to the plane ($\theta_D=1287$ K\cite{Tewary}) at least would have to double its value for double layer in order to explain our observed experimental results ($\theta_D=2110$ K). The perpendicular vibration modes do not change in such a substantial way because of the presence of a second graphene layer, and thus they do not cause the high increase on the displacement cross section. 

\begin{figure*}[t!]
 \subfigure[\ ]{\includegraphics[width=0.4 \textwidth]{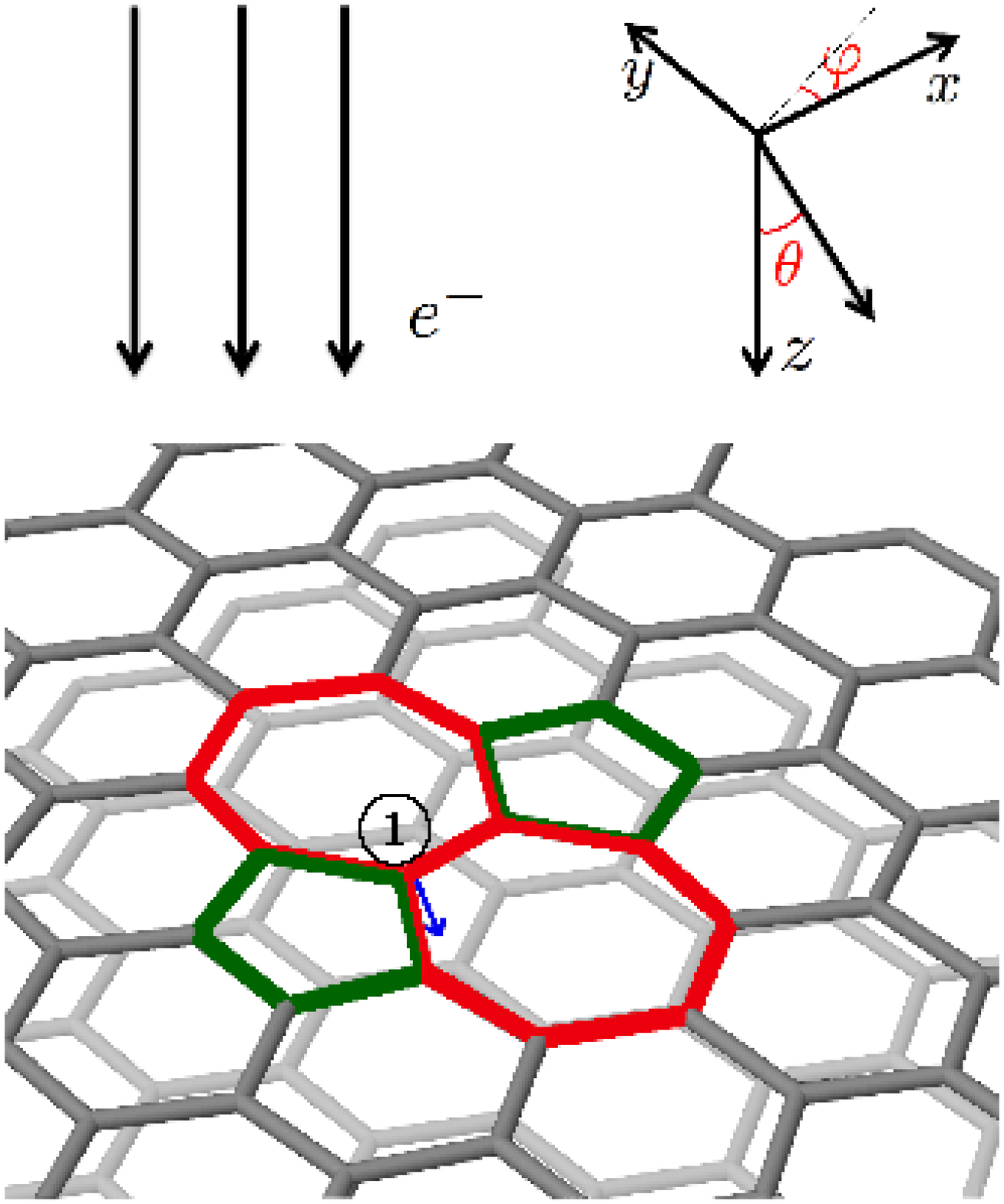}}
 \subfigure[\ ]{\includegraphics[width=0.45 \textwidth]{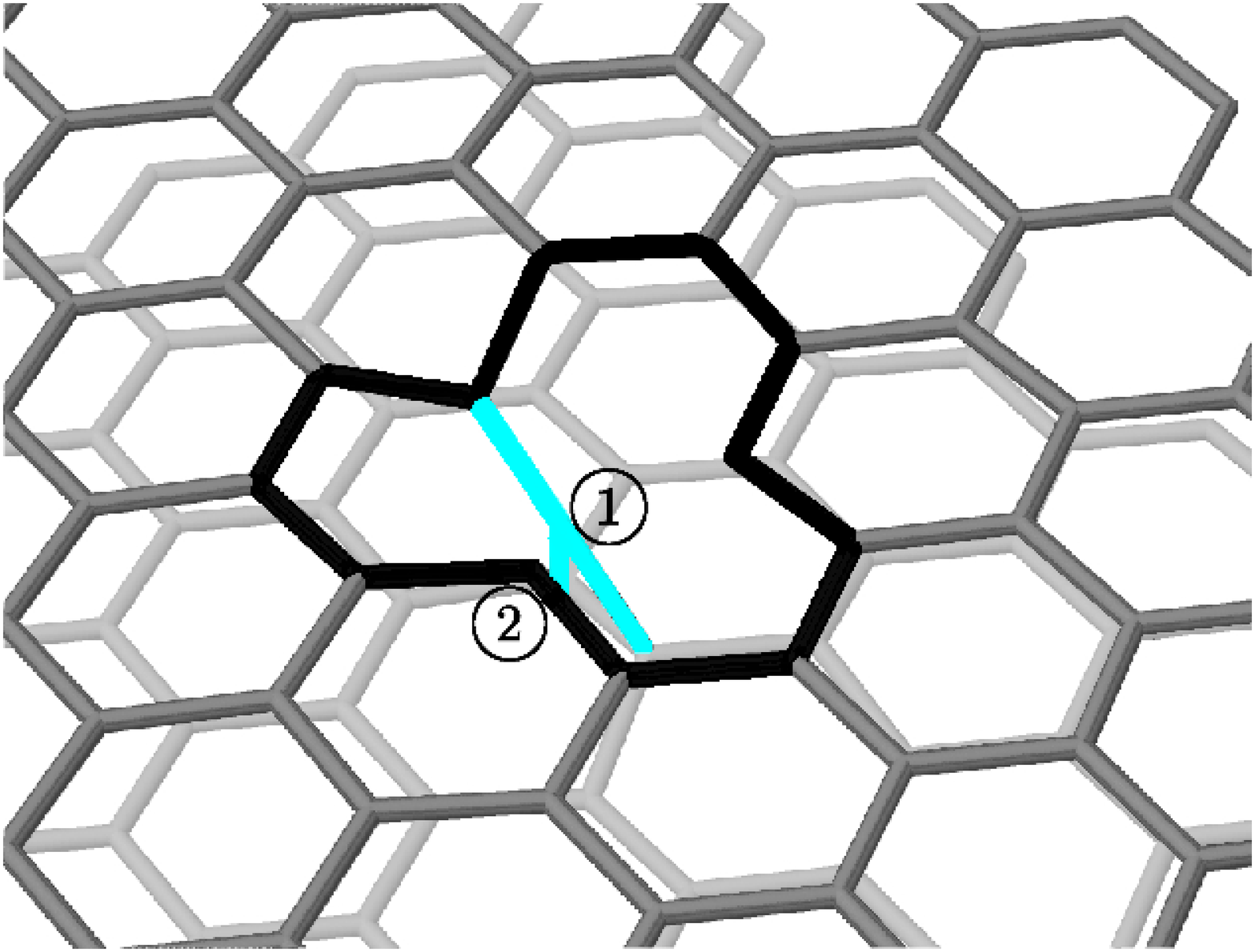}}
 \caption{(a) Initial state of the AIMD simulation with a Stone-Wales defect. It is made by two heptagonal (red) and two pentagonal (green) carbon rings. The blue arrow points in the direction ($\theta=12^{\circ}$, $\varphi=1^{\circ}$) of the initial velocity ($T=17$ eV) which is given to the scattered atom (atom 1). (b) Final state of the same simulation in which the intimate Frenkel pair is stabilized. In the upper layer a vacancy is formed (black) and the scattered atom remains trapped between both layers (blue, atom 1) bridging the upper layer (dark) with the lower layer (light). The atom 2 is `kicked' in the next simulation to obtain the intimate bi-Frenkel pair.}
\label{figura7}
\end{figure*}

Based on the previous results, we conclude that the creation process of vacancies is not caused by the direct expulsion of atoms. In addition, we observe that the lower layer has a similar kinetic behavior as the monolayer graphene system, which suggests that the origin of a different kinetic behavior comes from the collision of the electrons with the upper layer atoms.

\begin{table}[t!]
\caption{The expulsion threshold energy in terms of the energy acquired by the atom and the corresponding incoming electronic energy for each type of atom. A, B and C refer to different configurations of carbon atoms in bilayer graphene (see text).}
\centering
\begin{ruledtabular}
    \begin{tabular}{ c | c @{\hspace{0.5 cm}} c@{\hspace{0.5 cm}}  @{\hspace{0.5 cm}} c@{\hspace{0.5 cm}}  }
     & Type of atom & $E_{thr}($eV$)$&$E_{thr}^{el}($keV$)$ \\ \hline
    Monolayer &  & 22 & 110  \\
    Bilayer & A & 22 & 110  \\ 
     &B & 27 & 132  \\ 
     &C & 32-35 & 153-166  \\ 
    \end{tabular}
\end{ruledtabular}
\label{tabla3}
\end{table}

Once the possibility of defect formation due to direct ejection of atoms is discarded, we study the possibility of creation of intermediate states allowed by the presence of the second layer that would facilitate the formation of vacancies in the sample, i.e., a catalytic process. The most intuitive candidate that would play such a role is the Frenkel pair (Fig. \ref{figura7}), where the ``kicked'' atom does not escape from the system but remains trapped as an interstitial between both graphene layers, leaving behind a vacancy in its original position.\cite{Ewels} Telling \textit{et al.}\cite{Telling} have analyzed the energetics of the intimate Frenkel pair, conformed by an interstital atom neighboring a vacancy, and they have concluded that the stacking where this defect is more stable is the SP stacking, for which the formation energy and excess energy barrier are 10.6 eV and 1.4 eV, respectively. From these data, we can deduce the thermal activation barrier for the Pristine $\rightarrow$ intimate Frenkel process: $\Delta E_{p\rightarrow if}=12$ eV, and for the intimate Frenkel $\rightarrow$ Pristine process: $\Delta E_{{if\rightarrow p}}=1.4$ eV. We calculate the cross section related with each process by using Eqs. (\ref{eq.2}), (\ref{eq.3}), (\ref{eq.4}), and (\ref{eq.5}) and from Eq. (\ref{eq.10}) we obtain an approximate value of the relative concentration between carbon atoms and intimate Frenkel pairs in equilibrium:

\begin{equation}
\frac{[\text{C}]_e}{[\text{F}]_e}=\frac{\sigma_{{if\rightarrow p}}}{\sigma_{{p\rightarrow if}}}\approx\frac{34}{1}.
 \label{eq.15}
 \end{equation}
This last result indicates that approximately for each 34 carbon atoms, one intimate Frenkel pair should be in the sample. Consequently, this indicates that the population of this defect could be substantial enough to make it a good candidate for being the intermediate state for the vacancy formation process. It is rather counter-intuitive that such a difference in energy barriers (1.4 eV vs 12.0 eV) should give rise to that very large ratio of defects. It should be remembered though that these are not thermal processes but are related to the collision events, which transmit energies of several eV.

Our next step is to try to obtain a stable intimate Frenkel pair from an electron scattering event by AIMD below the expulsion threshold energy. Following the results obtained by Telling, \textit{et al.}\cite{Telling} we start our simulations from the SP stacking. We carry out the simulation for 17 different emission angles, and for each one we give seven different energies to the emitted atom within the range 16-22 eV. In all cases, the atom comes back to its original position and we never observe the stabilization of the intimate Frenkel pair. 

The reason for the apparent contradiction between the kinetic estimation in Eq. (\ref{eq.15}) and the explicit calculation of expulsion lies in the fact that we are distributing an excitation energy in a single atom, while the necessary energy for the most optimum pathway is estimated using thermal barriers and is produced when this energy is distributed in a particular way to several atoms. The AIMD results indicate that the direct Frenkel pair formation caused by single electron-atom collision is unlikely from a pristine graphene sample.

\subsubsection{Kinetics: multistep sputtering process}

Inasmuch as we have not been able to stabilize the intimate Frenkel pair below the expulsion threshold energy from a pristine sample, we study other possible intermediate states that could facilitate the formation of the intimate Frenkel pair. A previous study\cite{Ewels} has shown that this defect has two possible pathways for annihilation: it can be converted into the pristine configuration or into a Stone-Wales defect. This suggests that the Stone-Wales defect could be an intermediate step during the creation of the intimate Frenkel pair. The Stone-Wales defect is formed when a carbon-carbon bond is rotated $90^{\circ}$ creating two pentagonal and two heptagonal carbon rings (Fig. \ref{figura7}). This defect has lower formation energy than the vacancy and it has been already obtained by AIMD simulations from a pristine graphene sample below the expulsion threshold energy.\cite{Kotakoski} It is expected to form (and annihilate) under the irradiation in our experiment.

We first obtain via AIMD expulsion simulations the expulsion threshold energy for a monolayer graphene sample that contains the Stone-Wales defect, in order to establish the energetic limit: $E_{thr}^{SW}=18$ eV. Then, starting from a sample that contains one Stone-Wales defect we try to obtain the intimate Frenkel defect by AIMD simulations for lower energy values than this limit: after carrying out several simulations with different angles for the initial velocity and energies, we succeeded in stabilizing the intimate Frenkel pair. Figure \ref{figura7} shows the initial state of the simulation where the scattered atom is labeled (atom 1) and the final state in which the intimate Frenkel pair is stabilized. In this case, we used an energy of $T=17$ eV and an azimuthal angle $\theta=12^{\circ}$ and a polar angle $\varphi=1^{\circ}$ for the initial velocity of the scattered atom.

\begin{figure*}[t!]
\subfigure[\ ]{\includegraphics[width=0.45 \textwidth]{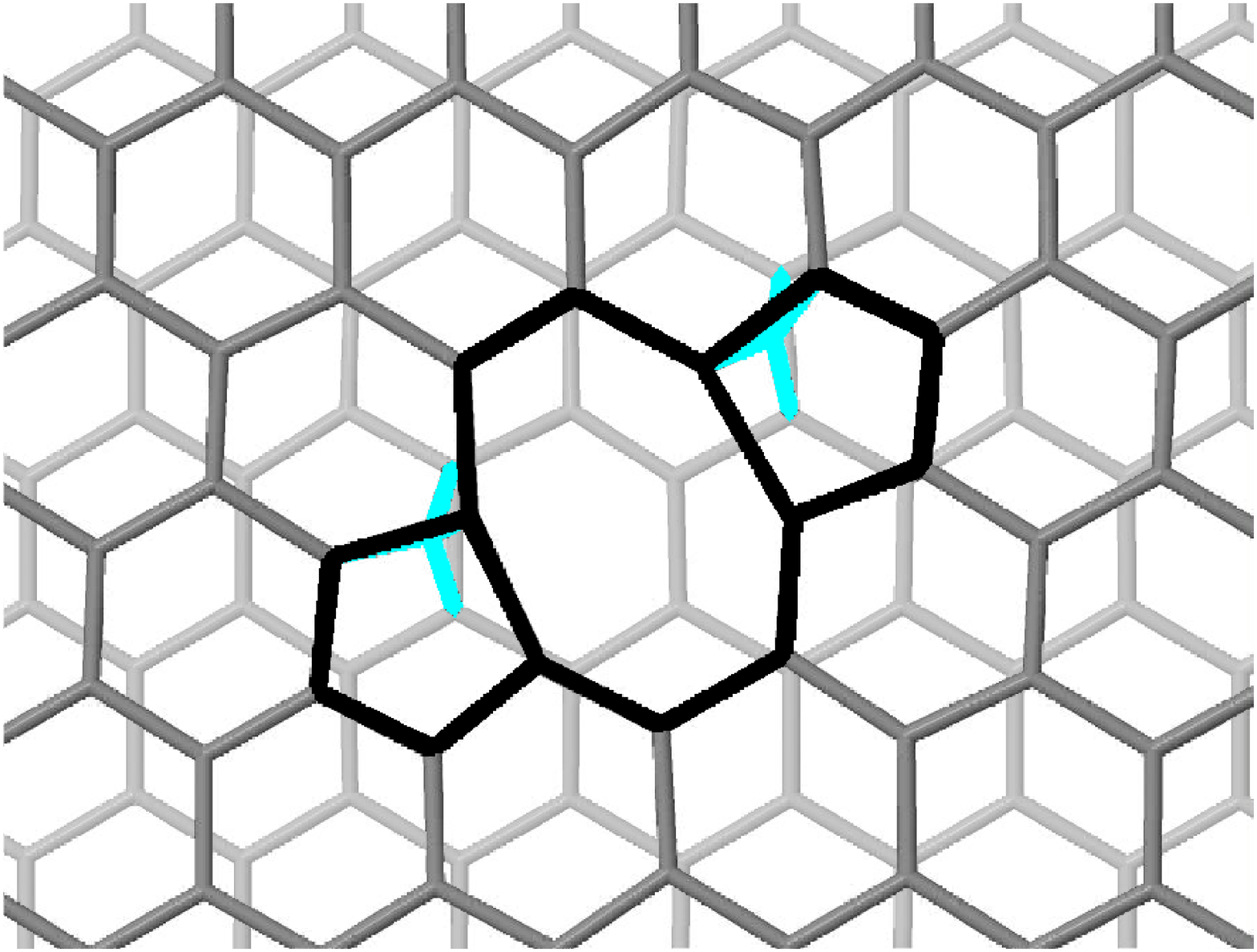}}
\subfigure[\ ]{\includegraphics[width=0.45 \textwidth]{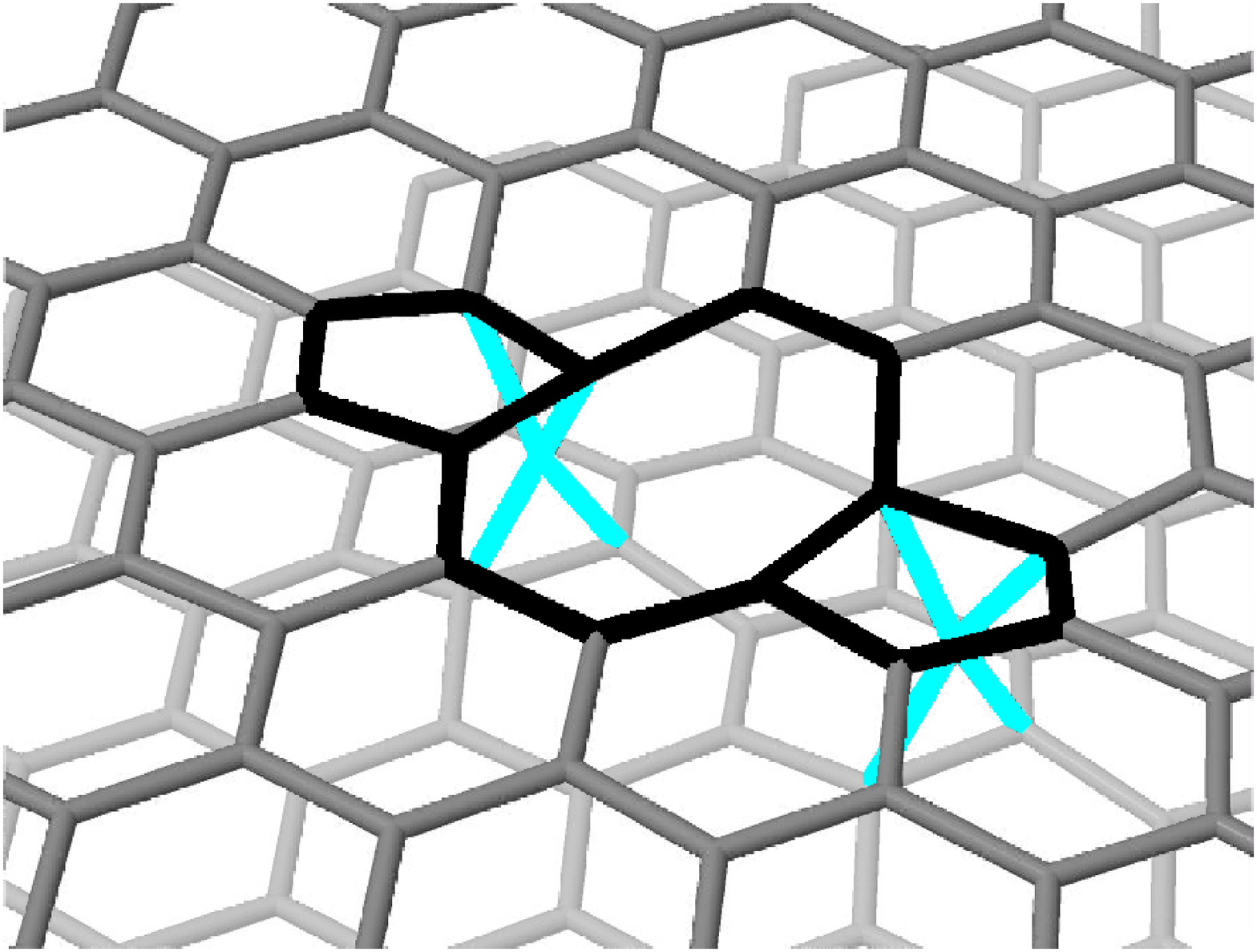}} 
\caption{Final relaxed configuration of an intimate bi-Frenkel pair. The initial state contains a unique intimate Frenkel pair (Fig. \ref{figura6}) and a certain velocity ($T=15$ eV, $\theta=20^{\circ}$, $\varphi=153^{\circ}$) is given to the scattered atom. The V$_2$(5-8-5) divacancy (black) is formed in the upper layer while the two scattered atoms (blue) are bridging the upper layer (dark) with the lower layer (light). (a) and (b) correspond to different perspectives.} 
\label{figura8}
\end{figure*}

Since vacancies or Frenkel pairs barely diffuse in the conditions of the experiment, and since the butterfly defect has two vacancies. we check the possibility of creating a second intimate Frenkel pair in the neighborhood of the previously formed one for lower energetic values than the expulsion threshold energy ($E_{thr}^{SW}$). For this purpose, we relax a bilayer graphene sample already containing an intimate Frenkel pair and we simulate a second collision event by giving a certain velocity to one of the atoms. We choose the atom labeled as 2 in Fig. \ref{figura7}(b) because it has a dangling bond, and therefore should be easier to expel from its original position. When this atom is `kicked' with an energy of $T=15$ eV, an azimuthal angle $\theta=20^{\circ}$ and a polar angle $\varphi=153^{\circ}$ for the initial velocity, two neighboring intimate Frenkel pairs are stabilized in the system and a V$_2$(5-8-5) divacancy is created in the upper layer. This defect, which we refer as the intimate bi-Frenkel defect, is formed by two spiro-interstitials (four-fold coordinated interstitial) neighboring a V$_2$(5-8-5) divacancy. Fig. \ref{figura8} shows two different perspectives of the obtained intimate bi-Frenkel defect, where the V$_2$(5-8-5) divacancy is located in the upper layer formed by two pentagonal and one octagonal carbon rings. This divacancy is the one experimentally observed before the butterfly defect stabilizes (see Fig. \ref{figura5}). We obtain a formation energy of 15.54 eV for the obtained intimate bi-Frenkel pair.

 \begin{table}[t]
\caption{Activation energies corresponding to each process: The first four values are thermal activation barriers while the last one has been obtained by our AIMD simulations.}
\centering
\begin{ruledtabular}
    \begin{tabular}{l @{\hspace{0.1 cm}} c @{\hspace{0.4 cm}}}
    Process & Activation \\ 
    & energy (eV) \\ \hline
    Pristine $\longrightarrow$ Stone-Wales (Ref. 7) & 9  \\ 
    Stone-Wales $\longrightarrow$ Pristine (Ref. 7) & 5.5  \\ 
    Stone-Wales $\longrightarrow$ I. Frenkel (Refs. 7 and 44) & 8.5  \\ 
    I. Frenkel $\longrightarrow$ Pristine (Refs. 7 and 44) & 1.4  \\ 
    I. Frenkel $\longrightarrow$ I. Bi-Frenkel & 15 \\
    \end{tabular}
 \end{ruledtabular}
    \label{tabla4}
    \end{table}

By AIMD simulations we have demonstrated the possibility of creating divacancies in bilayer graphene below the expulsion threshold energy by means of catalyzed intermediate states. To measure the frequency in which the formation of divacancies occurs following the suggested pathway, we propose the following chain of reactions:

\begin{displaymath}
\xymatrixcolsep{3pc}
\xymatrix{
  \mbox{Pristine} \ar[r] & \mbox{Stone-Wales} \ar[l] \ar[dl] \\ 
  \mbox{I. Frenkel} \ar[u] \ar[r] & \mbox{I. bi-Frenkel}
}
\end{displaymath}
The intimate bi-Frenkel defect does not easily annihilate back into single vacancies because all the atoms have fully satisfied bonds. This is corroborated by the experiment, where, once a divacancy is created, it is completely stable. By using the activation energies shown in Table \ref{tabla4} for each possible reaction and the Eqs. (\ref{eq.2}), (\ref{eq.3}), (\ref{eq.4}), and (\ref{eq.5}) we calculate the cross section of each possible reaction. Solving the first order rate equations of this chain of reactions (see Appendix), we obtain a theoretical estimation of 195 intimate bi-Frenkel defects that should have been formed in our sample or a corresponding value for the cross section of $\sigma=75.2$ mb. This last value is overestimated comparing it with the obtained experimental one. The overestimation of the cross section was already expected by the use of thermal activation barriers and isotropic cross sections. The order of magnitude is, however, correct, showing that the proposed process is consistent with the experiments.

\section{Conclusions}

We have observed that the cross section for the formation of butterfly defects in bilayer graphene under electron radiation is substantially higher than in the case of monolayer graphene. Another difference with respect to the monolayer case is that there are no different types of defect in the sample, but only one type of divacancy: the butterfly defect, which stabilizes within the SP and AB stackings of the Moir\'e pattern becoming a very stable defect. Although its creation is facilitated by the presence of a second graphene layer, the filtering of the image shows that it is located in only one of the layers. 

We find that the stability of the butterfly defect does not change significantly between monolayer and bilayer graphene, and the atomic displacements are small. These results are consistent with the weak interaction between two graphene layers and confirms that the mechanism for the formation process of the butterfly defect is different in the monolayer and bilayer graphene cases.

The results of the expulsion threshold energies for different atoms in bilayer graphene shows that the ones located in the lower layer (farther away from the beam source) have a similar kinetic behavior to the ones from monolayer graphene, hence the layer closer to the beam is the one that contains the divacancies.

We have demonstrated the possibility of creating divacancies in a bilayer graphene sample by AIMD simulations for electronic energies that are below the expulsion threshold energy. This is possible because new intermediate catalyzed states are created due to the presence of the second graphene layer. Although the estimated concentration of divacancies formed following the suggested chain of reaction is overestimated, the order of magnitude is correct with experimental results. Therefore, we demonstrate the principal possibility of creating vacancies in a multilayer graphitic sample with lower electronic energies than the expulsion threshold, and accordingly, an increase of the displacement cross section of such systems with respect to the monolayer graphene case. 

The reason why the divacancies stabilize within the SP and AB stackings of the Moir\'e pattern still remains unclear. However, as an initial proposal, we think that the fact that the interstitial atoms are stabilized in the SP stacking,\cite{Telling2,Telling,Ewels} catalyze the formation of Stone-Wales defects,\cite{Ewels2} and thus facilitate the proposed chain of reactions, could be related with the unresolved part of the observed phenomenon. 

\section{Acknowledgments}
SGIker (UPV/EHU, MICINN, GV/EJ, ERDF and ESF) support is gratefully acknowledged. The calculations were performed on the following HPC clusters: Tortilla (CIC nanoGUNE, Spain) and Arina (Universidad del Pais Vasco/Euskal Herriko Unibertsitatea, Spain).

\appendix
\section{Calculation of intimate bi-Frenkel population}
The proposed chain of reactions is summarized in the next diagram:
\\

\begin{displaymath}
\xymatrixrowsep{5pc}
\xymatrixcolsep{7pc}
\xymatrix{
  \mbox{P} \ar[r]_{\mbox{$k$}_{\mbox{\scriptsize{SW-P}}}} & \mbox{SW} \ar[l]_{\mbox{$k$}_{\mbox{\scriptsize{P-SW}}}} \ar[dl]^{\mbox{$k$}_{\mbox{\scriptsize{SW-IF}}}} \\ 
  \mbox{IF} \ar[u]^{\mbox{$k$}_{\mbox{\scriptsize{IF-P}}}} \ar[r]_{\mbox{$k$}_{\mbox{\scriptsize{IF-IBF}}}} & \mbox{IBF}
}
\end{displaymath}
where each species (Pristine graphene, Stone-Wales, intimate Frenkel and intimate bi-Frenkel) is represented by its initials and each reaction is characterized by a rate constant $k$. Assuming that these reactions follow a first order rate law, the time-dependent concentration of each species is obtained by solving the following system of differential Eqs.:
\begin{align}
\dfrac{d[\text{P}]}{dt}&=-\mbox{$k$}_{\mbox{\scriptsize{P-SW}}}[\text{P}]+\mbox{$k$}_{\mbox{\scriptsize{SW-P}}}[\text{SW}]+\mbox{$k$}_{\mbox{\scriptsize{IF-P}}}[\text{IF}],\\
\dfrac{d[\text{SW}]}{dt}&=-(\mbox{$k$}_{\mbox{\scriptsize{SW-P}}}+\mbox{$k$}_{\mbox{\scriptsize{SW-IF}}})[\text{SW}]+\mbox{$k$}_{\mbox{\scriptsize{P-SW}}}[\text{P}],\\
\dfrac{d[\text{IF}]}{dt}&=-(\mbox{$k$}_{\mbox{\scriptsize{IF-P}}}+\mbox{$k$}_{\mbox{\scriptsize{IF-IBF}}})[\text{IF}]+\mbox{$k$}_{\mbox{\scriptsize{SW-IF}}}[\text{SW}],\\
\dfrac{d[\text{IBF}]}{dt}&=\mbox{$k$}_{\mbox{\scriptsize{IF-IBF}}}[\text{IF}],
\end{align}
together with a boundary condition for the initial concentration of each species: $[\text{P}]_0=38.46$ nm$^{-2}$, $[\text{SW}]_0=[\text{IF}]_0=[\text{IBF}]_0=0$. Once the $[\text{IBF}]$ function is calculated, it is evaluated at the time value related with the employed electronic dose in the experiment.

\bibliography{papermodel}

{}
  
\end{document}